\documentclass[a4paper,fleqn,usenatbib]{mnras}

\usepackage{newtxtext,newtxmath}

\usepackage[T1]{fontenc}
\usepackage{ae,aecompl}

\usepackage{graphicx}	
\usepackage{amsmath}	
\usepackage{amssymb}	

\title[Diversity of $f_{\rm esc}$ and its origins]{Diversity of the Lyman continuum escape fractions of high-$z$ galaxies and its origins}

\author[T. Sumida et al.]{
Takumi Sumida,$^{1}$\thanks{E-mail: sumida.takumi@j.mbox.nagoya-u.ac.jp}
Daichi Kashino,$^{2}$\thanks{E-mail: kashinod@phys.ethz.ch}
Kenji Hasegawa,$^{1}$\thanks{E-mail: hasegawa.kenji@a.mbox.nagoya-u.ac.jp}
\\
$^{1}$Division of Particle and Astrophysical Science, Graduate School of Science, Nagoya University, Nagoya, 464-8602, Japan\\
$^{2}$Department of Physics, ETH Z{\"u}rich, Wolfgang-Pauli-strasse 27, CH-8093 Z{\"u}rich, Switzerland
}

\date{Accepted XXX. Received YYY; in original form ZZZ}

\pubyear{2016}

\begin{document}
\label{firstpage}
\pagerange{\pageref{firstpage}--\pageref{lastpage}}
\maketitle

\begin{abstract}
The Lyman continuum (LyC) escape fraction is a key quantity to determine the contribution of galaxies to cosmic reionization. 
It has been known that the escape fractions estimated by observations and numerical simulations show a large diversity. 
However, the origins of the diversity are still uncertain. 
In this work, to understand what quantities of galaxies are responsible for controlling the escape fraction, we numerically evaluate the escape fraction by performing ray-tracing calculation with simplified disc galaxy models. 
With a smooth disc model, we explore the dependence of the escape fraction on the disposition of ionizing sources, and find that the escape fraction varies up to $\sim 3$ orders of magnitude. 
It is also found that the halo mass dependence of disc scale height determines whether the escape fraction increases or decreases with halo mass. 
With a clumpy disc model, it turns out that the escape fraction increases as the clump mass fraction increases  because the density in the inter-clump region decreases. 
In addition, we find that clumpiness regulates the escape fraction via two ways when the total clump mass dominates the total gas mass; the escape fraction is controlled by the covering factor of clumps if the clumps are dense sufficient to block LyC photons, otherwise the clumpiness works to reduce the escape fraction by increasing the total number of recombination events in a galaxy. 
\end{abstract}

\begin{keywords}
cosmology: dark ages, reionization, first stars - galaxies: high-redshift - galaxies: structure
\end{keywords}



%
%

\section{INTRODUCTION}\label{INTRO}
Recently, observations have successfully revealed the outline of cosmic reionization \citep[e.g.,][]{2006AJ....132..117F,2011ApJS..192...18K,2016A&A...596A.108P}. 
The epoch of reionization (EoR) is the rendezvous point where theoretical studies on the first generation objects meet  observational studies on distant galaxies. 
Although spectra of high-$z$ quasars are indicating that reionization completed by $z\sim6$ \citep{2006AJ....132..117F}, the detailed process of reionization is still highly unknown.  
One of unresolved factors for understanding the detailed process of reionization is the type of ionizing sources. 

The first stellar objects formed in the Universe are theoretically thought to be massive \cite[e.g.][]{Bromm02, Abel02, Yoshida08, Hirano14, Susa14}. 
They and their remnants are often expected to contribute reionization \cite[e.g.][]{2000MNRAS.314..611C, 2004ApJ...604..484M, Sokasian04, Ahn12}. 
However there exist no observational constraints on the amount of their contribution up to date.

Luminous active galactic nuclei (AGNs), such as quasars, have traditionally been considered to be minor contributors to reionization due to the rapid decrease in their abundance at $z>3$ \cite[e.g.][]{Masters12,Ueda14}. 
However, if faint AGNs are abundant as much as \citet{Giallongo15} indicated at $z>6$, they could have a potential to ionize the Universe \cite[e.q.][]{MadauHaardt15,Khaire16,Mitra16,Yoshiura17}, although the abundance of faint AGNs during the EoR remains unclear. 

Since a number of galaxies, such as Lyman $\alpha$ emitters and Lyman break galaxies, have already been observed \cite[e.g.][]{Iye06,Kashikawa06,Oesch09,Ouchi09,Shibuya12,Konno14,Bouwens15,Santos16}, some studies have insisted that galaxies had the largest contribution to reionization \citep[e.g.,][]{2011MNRAS.410..775R,2012MNRAS.425.1413F,2015ApJ...802L..19R,2015ApJ...799...12I}. 
In fact, most numerical simulations of cosmic reionization have assumed galaxies as  standard ionizing sources \cite[e.g.][]{Iliev06,Trac07,Baek09,Hasegawa13}. 
The contribution of galaxies to reionization is largely controlled by three factors, all of which are, however, still remain poorly constrained.
The first one is the abundance of faint galaxies \citep{2015ApJ...802L..19R, 2015ApJ...799...12I}.
Although numerical simulations have predicted that a number of faint galaxies with absolute UV magnitude being $M_{\rm UV}>-17$ exist during the EoR \citep[e.g.,][]{Finlator11,Wise14}, observational constraints on the abundance of such faint galaxies still have a large uncertainty. 
Next generation telescopes such as {\it James Webb Space Telescope}, {\it European Extremely Large Telescope} and {\it Thirty Meter Telescope} are expected to detect those faint galaxies.  
The second unknown factor is the production rate of Lyman continuum photons per unit star formation rate.  
It largely depends on the metallicity and the initial mass function of stars \cite[e.g.][]{2002A&A...382...28S,2015ApJ...800...97T}.
The last one is the LyC escape fraction ($f_{\rm esc}$), which indicates the fraction of the number of LyC photons escaping from a galaxy to the number of LyC photons intrinsically produced in the galaxy. 
It has been demonstrated that the average escape fraction of $\sim20\%$ enables galaxies alone to  reionize the Universe \citep{2011MNRAS.410..775R,2012MNRAS.425.1413F,2015ApJ...802L..19R}, although the critical value depends somewhat on the number of faint galaxies and the intrinsic LyC photon production rate.

Observations have measured escape fractions at $z\sim3$, where $f_{\rm esc}$ is known to vary largely from one galaxy to another \citep{2006ApJ...651..688S,2006MNRAS.371L...1I,2009ApJ...692.1287I}. 
In particular, \citet{2006A&A...448..513B} have suggested that clumpy distribution of neutral hydrogen gas in galaxies as one of origins of the large diversity in $f_{\rm esc}$.
At higher redshifts ($z>3$), most LyC photons that escape from galaxies are immediately absorbed by the intergalactic medium (IGM). 
Therefore, it is almost impossible to directly detect LyC photons from the EoR even with the next generation telescopes. 

Because of the difficulty of direct detection of LyC radiation, theoretical approaches are essentially important for revealing $f_{\rm esc}$ during the EoR. 
Theoretical estimates of $f_{\rm esc}$ by numerical simulations have been actively performed \citep[e.g.,][]{2008ApJ...672..765G,2010ApJ...710.1239R,2011MNRAS.412..411Y,Wise14,Paardekooper15,Kimm17}. 
However their results are not consistent with each other. 
\citet{2008ApJ...672..765G} have firstly investigated the escape fraction of galaxies in a cosmological hydrodynamics simulation and found that $f_{\rm esc}$ increases with increasing halo mass, while most other studies have shown the opposite trend, i.e., $f_{\rm esc}$ decreases as halo mass increases. 
Even if we focus on the simulations indicating the latter trend, the average escape fractions of halos with a given mass scale are considerably different for each work. 
In addition, studies with a large sample of simulated galaxies \cite[e.g.][]{2011MNRAS.412..411Y} have also found that $f_{\rm esc}$ shows very large diversity of the escape fractions for fixed halo mass. 
In summary, while the escape fractions derived from previous simulations seem to show various values and trends, we still do not have enough understanding on their origins due to  different characteristics of each simulation (spatial resolution, numerical method, implemented physics and so on) as well as the complicated properties of galaxies in state-of-the-art cosmological hydrodynamic simulations. 

In contrast to studies with such ``realistic'' galaxies that are suitable for predicting  quantitative values of $f_{\rm esc}$, studies with simplified galaxy models are often useful to understand the essence of the escaping process of LyC photons \cite[e.g.][]{1994ApJ...430..222D,2000ApJ...531..846D,2002MNRAS.337.1299C,2011ApJ...731...20F}. 
For instance, \citet{2011ApJ...731...20F} (hereafter FS11) widely investigated the impact of clumps on $f_{\rm esc}$ by utilizing a simple disc galaxy model, assuming only one ionizing source located at the centre of a galaxy. 
They have found that the escape fraction is very sensitive to the properties of clumps, such as volume filling factor, and their internal gas density. 
\cite{Umemura12} have also pointed out that inhomogeneous gas distribution in a galaxy affects the escape fraction. 
Although these studies have not evaluated the dependence of $f_{\rm esc}$ on halo mass, it is highly expected that the gas distribution in a galaxy is one of important factors for determining the escape fraction. 
Observations have actually indicated that galaxies have clumpy structures at low redshifts $z<3$ \citep[e.g.,][]{2010MNRAS.406..535P,2010ApJ...710..979O,2012ApJ...757..120G,2014ApJ...786...15M}.  
In addition, \citet{2016ApJ...821...72S} have found that the number fraction of clumpy galaxies  depends on redshift; the number fraction increases with increasing redshift at $1<z<3$, while it decreases as redshift increases at $3<z<8$. 
This fact may imply the redshift evolution of $f_{\rm esc}$. 
Furthermore from a theoretical point view, \citet{2016MNRAS.456.2052I} have performed a Toomre analysis of high redshift galaxies in cosmological simulations and found that massive clumps are formed in the galaxies. 

In this paper, we aim to understand what properties of galaxies control the LyC escape fraction and cause large diversity in $f_{\rm esc}$ as seen in previous studies.
In order to clarify the essential quantities that determine the escape fraction, we dare to employ a simplified galaxy model, which is an extended version of the one used in FS11.  
We investigate the effects of multiple ionizing sources located elsewhere in a disc and the halo mass dependence, which are not considered in FS11. 
However, one may feel our model is still too simple to predict $f_{\rm esc}$ quantitatively. 
Indeed, we do not aim to predict it, but to understand which properties of galaxies impact the escape fraction. 

This paper is organized as follows. 
In Section \ref{sec:METHODOLOGY}, we explain the setup of our numerical calculation. 
In Section \ref{sec:RESULTS}, we show the results of calculation and discuss the implications of our results. 
Finally in Section \ref{sec:SUMMARY} we summarize our conclusions. 
Throughout this paper, we use the cosmological parameters from \citet{2016A&A...594A..13P}: $\Omega_{\rm m0}=0.31$, $\Omega_{\rm b}=0.048$, $\Omega_{\rm \Lambda}=0.69$ and $h=0.68$. 

%
%

\begin{figure*}
\begin{center}
\includegraphics[width=0.4\textwidth]{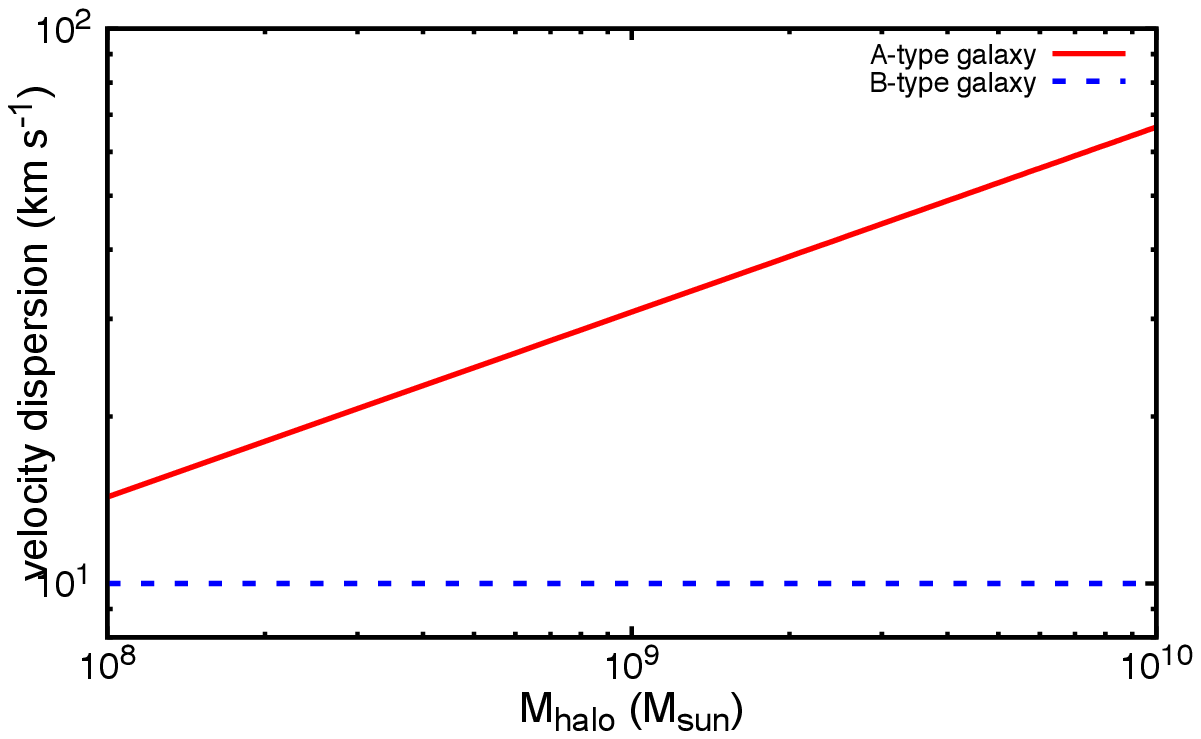}
\includegraphics[width=0.4\textwidth]{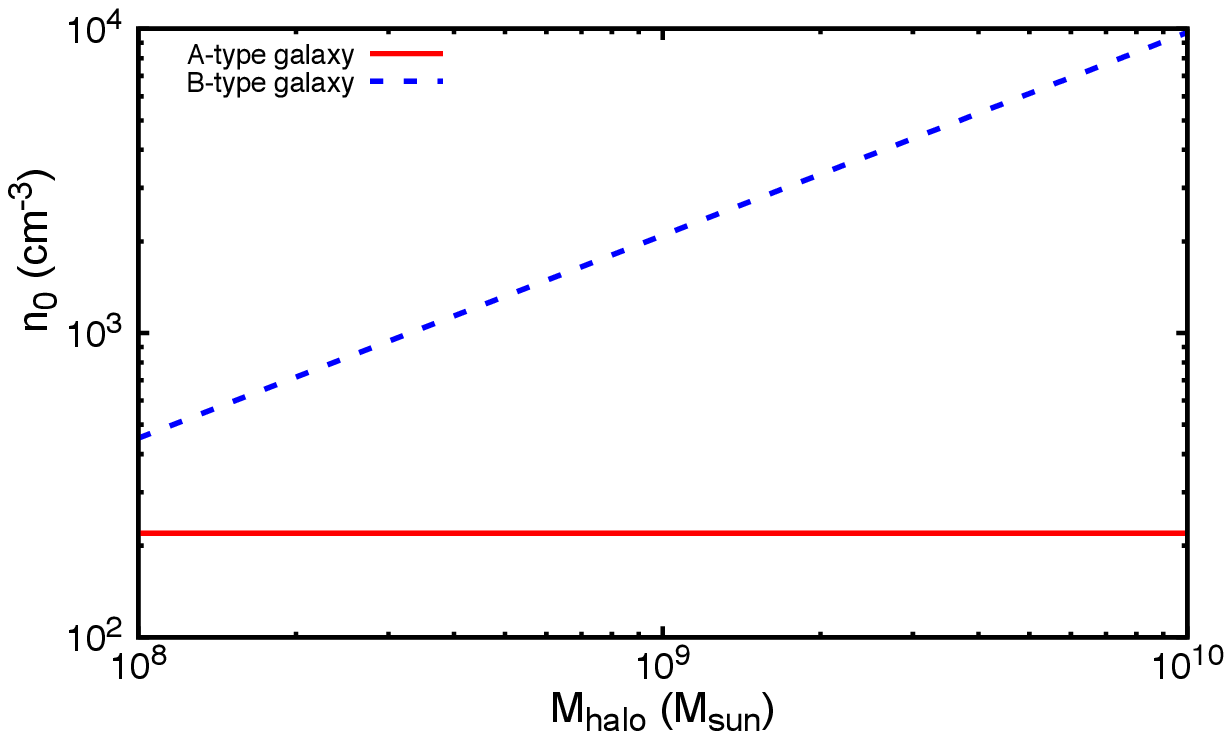}
\includegraphics[width=0.4\textwidth]{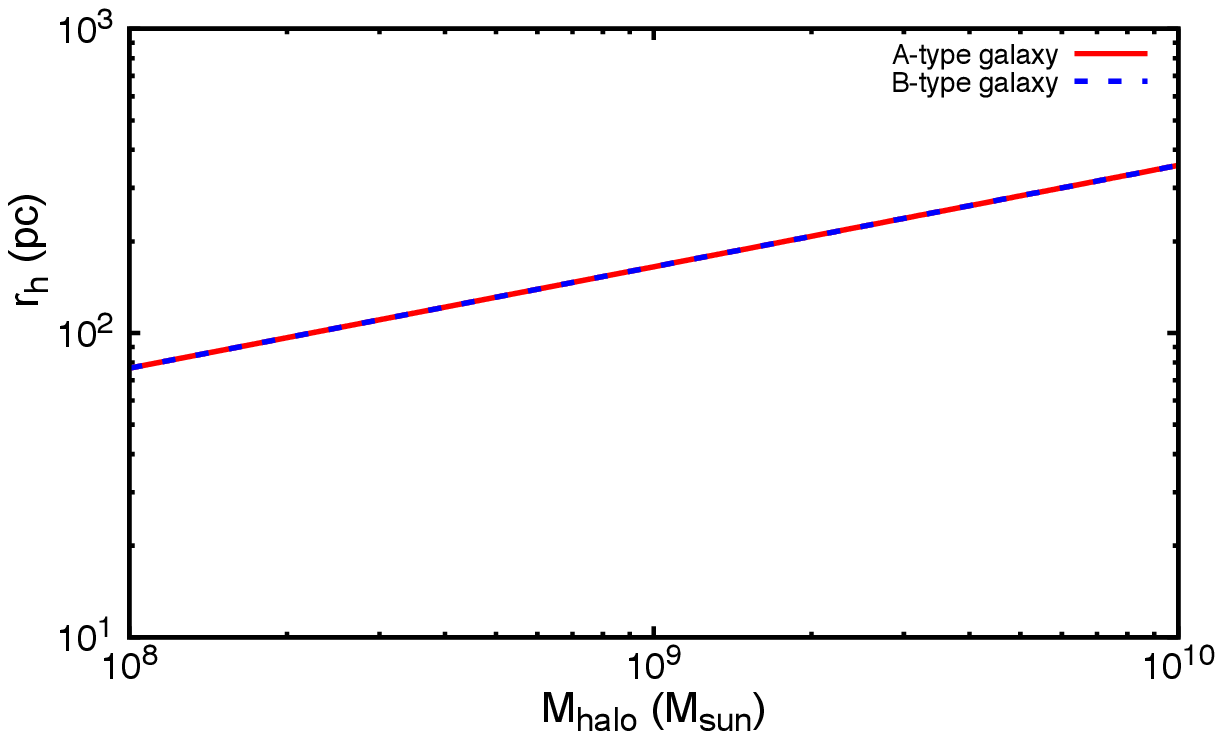}
\includegraphics[width=0.4\textwidth]{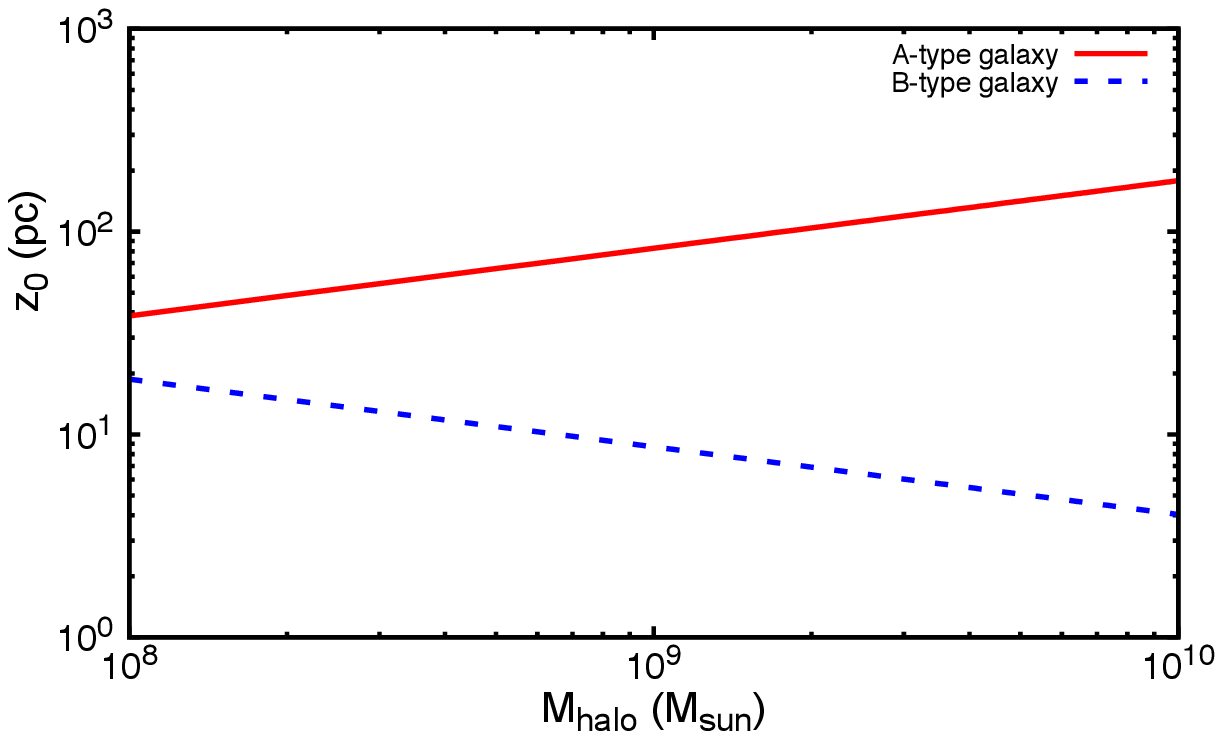}
   \caption{Velocity dispersion (upper left), scale radius (lower light), central hydrogen number density (upper right) and scale height (lower right)  as a function of halo mass.
   In each panel, the red solid and blue dashed lines respectively show type-A galaxies (i.e., $\langle v^2 \rangle =GM_{\rm halo}/r_{\rm vir}$) and type-B galaxies (i.e., $\sqrt{\langle v^2 \rangle}=10~{\rm km~s^{-1}}$).}
\end{center}
   \label{fig:galala}
\end{figure*}

\section{METHODOLOGY}\label{sec:METHODOLOGY}
In this section, we describe the setup of our numerical calculation to evaluate the escape fraction.
We employ galaxy models similar to that in FS11, expect for the following three points.
First, we consider two types of disk models that have different dependence of the disc scale hight on the halo mass. 
We describe in detail how the results depend on the disc models in \S \ref{mdepend_s}.
Second, our parametrization for characterizing clumps is modified from that by FS11 (\S \ref{sec:Galactic Geometry}). 
Our parameterization allows us to more clearly understand the behavior of the escape fraction (\S\ref{sec:clumps}).  
Third, we consider multiple ionizing sources whereas FS11 consider only one ionizing source residing at the galactic center. 

\subsection{Model of galaxy}\label{sec:Galactic Geometry}
First let us describe our galaxy model. 
We consider a galaxy that forms within a cold dark matter halo that follows an NFW density profile \citep{1997ApJ...490..493N}.  
The virial radius of a halo with the halo mass $M_{\rm halo}$ that formed at $z=z_{\rm f}$ is given as
\begin{align}
	r_{\rm vir} = 0.76\left(\frac{M_{\rm halo}}{10^8{\rm M_{\odot}}h^{-1}} 
	\right)^{1/3}\left(\frac{\Omega_{\rm m0}}{\Omega_{\rm m} 
	(z_{\rm f})}\frac{\Delta_{\rm c}}{200} \right)^{-1/3}\nonumber\\
	\times \left(\frac{10}{1+z_{\rm f}} \right)h^{-1} \ \ {\rm kpc},
\end{align}
where $h$ is the hubble parameter, $\Omega_{\rm m0}$ and $\Omega_{\rm m}(z_{\rm f})$ are respectively the cosmological density parameters at the present-day and the formation epoch $z_{\rm f}$ for which we assume $z_{\rm f}=6$. 
The threshold overdensity for a virialized dark matter halo is defined as $\Delta_{\rm c} = 18\pi^2+82d-39d^2$,where $d = \Omega_{\rm m}(z_{\rm f})-1$ \citep{1998ApJ...495...80B}. 
We assume a disc-like structure purely composed of hydrogen whose number density profile given as
\begin{equation}
	n_{\rm H}(r,Z)=n_0\ {\rm exp}[-r/r_{\rm h}]\ {\rm sech}^2 \left(\frac{Z}{z_0}\right),
\end{equation}
where $r$ is the radial distance, $Z$ is the height above the galaxy mid-plane, $n_0$ is the hydrogen number density at the centre of the galaxy, $r_{\rm h}$ is the scale radius, and $z_0$ is the scale height \citep{1942ApJ....95..329S}. The scale radius is given by
\begin{equation}
	r_{\rm h} = \frac{j_{\rm d}\lambda}{\sqrt{2}m_{\rm d}}r_{\rm vir},
\end{equation}
where $j_{\rm d}$ is the fraction of the angular momentum of the baryonic disc to the total angular momentum of the halo, $\lambda$ is the spin parameter, and $m_{\rm d}$ is the mass fraction of the disc to the halo. We adopt $j_{\rm d}/m_{\rm d}=1$ and $\lambda=0.05$ as in previous studies \cite[e.g.][]{2000ApJ...545...86W}. 
The scale height is given by
\begin{equation}
	z_0=\left(\frac{\langle v^2\rangle}{2\pi G\rho_0}\right)^{1/2},
\end{equation}
where $\langle v^2\rangle$ is the velocity dispersion and $\rho_0$ is the central mass density of the halo. 
As mentioned above, we consider two galaxy models: 1) a system with $\langle v^2 \rangle =GM_{\rm halo}/r_{\rm vir}$ (hereafter type-A galaxy) or 2) $\sqrt{\langle v^2 \rangle}=10~{\rm km~s^{-1}}$ (hereafter type-B galaxy). 
FS11 have employed the former model in which the scale hight is determined by the virial temperature. 
As usually expected, a gaseous disc becomes thinner than this model if radiative cooling effectively works. 
The balance between photo-heating and radiative cooling keeps the temperature of ionized gas being  
$\sim 10^4$~K \citep{TW96}. 
Therefore, in the latter model, the velocity dispersion is set to be the sound speed with 
$\sim 10^4$~K to represent an extreme case that the gas temperature does not obey the virial temperature. 
Note that real galaxies likely show more complicated disc structures, but these simplified models are very useful to understand how the geometry of the discs affects the escape fraction.

\begin{figure*}
\includegraphics[angle=-90,width=\columnwidth]{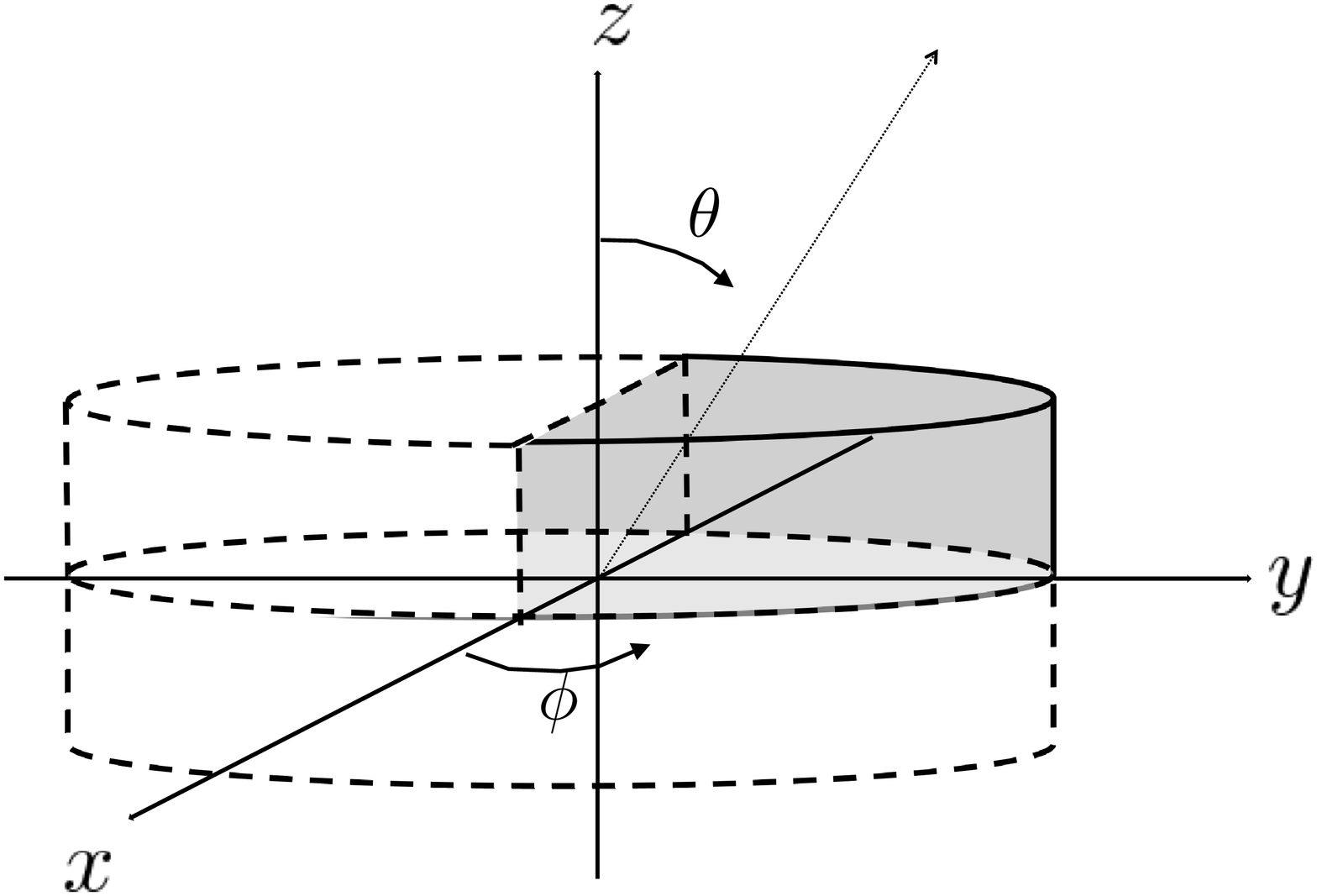}
	\caption{Schematic view of our simulation domain. The region surrounded by the 
	dashed curve corresponds to the whole disc, while the shaded region is the representative 
	domain in which numerical calculations are conducted. 
	The dotted arrow indicates a sightline discussed in \S~\ref{sightline}}
	\label{fig:schematic}
\end{figure*}
We fix the maximum radius and height of a disc to be $10r_{\rm h}$ from the galactic center and $5z_0$ from the galactic plane, respectively. 
The central mass density is given by,
\begin{align}
	\rho_0 = m_{\rm H}n_0 = 
 \frac{M_{\rm gas}}{2\pi\int{\rm exp}[-r/r_{\rm h}]{\rm sech}^2(Z/z_0)r{\rm d}r{\rm d}Z},
 \label{eq:rho_cal}
\end{align}
where $M_{\rm gas} = M_{\rm disc}-M_{\ast}$ is the galactic gas mass. 
Here, $M_{\rm disc} = m_{\rm d}M_{\rm halo}$ is the mass of the disc and $M_{\ast}$ is the stellar mass of the galaxy defined as $M_{\ast}=f_{\ast}M_{\rm disc}$ with $f_{\ast}$ being the stellar mass fraction. 
The disc mass fraction $m_{\rm d}$ is set to be the cosmic mean baryon fraction, i.e., $m_{\rm d}=\Omega_{\rm b}/\Omega_{\rm m}=0.155$.

Figure~\ref{fig:galala} shows the velocity dispersion, scale radius, scale height and central hydrogen number density as a functions halo mass for the two types of galaxies.
As for a type-A galaxy with $M_{\rm halo} =10^9~\rm M_{\odot}$, the scale radius, the scale height and the velocity dispersion are $165$ {\rm pc}, $82.8\ {\rm pc}$ and $\sqrt{\langle v^2 \rangle}=30~{\rm km~s^{-1}}$, respectively. 
On the other hand, as for a type-B galaxy with the same halo mass, these quantities are respectively $165$ {\rm pc}, $8.69\ {\rm pc}$ and $\sqrt{\langle v^2 \rangle}=10~{\rm km~s^{-1}}$. 
For type-A galaxies, we find that both the scale radius and the scale height are proportional to $M_{\rm halo}^{1/3}$, while the central hydrogen number density is independent of halo mass. 
On the other hand, for type-B galaxies, the scale height is found to be proportional to $M_{\rm halo}^{-1/3}$, because the central hydrogen number density depends on the halo mass as $n_0\propto M_{\rm halo}^{2/3}$.

In addition to the smooth disc component, we also consider clumps characterized by three parameters that are different from those in FS11; 
$n_{\rm c}$ the hydrogen number density in a clump, $N_{\rm c}$ the number of clumps, and $R_{\rm c}$ the diameter of a clump. 
The hydrogen number density in the inter-clump region (i.e., the smooth component) is given by 
\begin{equation}
	n_{\rm ic}(r,Z)=n'_0\ {\rm exp}[-r/r_{\rm h}]\ {\rm sech}^2 \left(\frac{Z}{z_0}\right),
	\label{eq:inter_eq}
\end{equation}
where $n'_0$ is the hydrogen number density at the centre of a galaxy, which is written as, 
\begin{equation}
	n'_0 = (1 - f_{\rm c,m})n_0.
\label{eq:rechange_no}
\end{equation}
Here $f_{\rm c,m}$ is the clump mass fraction defined as the fraction of the total clump mass $M_{\rm clump}$ to the gas mass in a disc $M_{\rm gas}$, where the total clump mass is described as 
\begin{equation}
	M_{\rm clump} = \frac{4\pi}{3} m_{\rm H} n_{\rm c} N_{\rm c} \left(\frac{R_{\rm c}}{2} \right)^3.
\label{eq:N_c}
\end{equation} 
We keep the scale height and scale radius of the smooth gas disc unchanged even if we include clumps. 

Let us briefly describe what we expect in the case that clumps exist. 
Since the recombination rate is proportional to the square of the number density, clumps often work to prevent LyC photons from escaping \citep[so-called shadowing effect, e.g.,][]{1999ASPC..181..227H}.  
A clump can completely block LyC photons if the diameter of the clump is greater than the $\rm Str\ddot{o}mgren$ length \citep{radtr1} which is defined as  
\begin{equation}
	l_{\rm s} = \frac{Q_{\rm LyC}}{4\pi r_\ast^2 \alpha_{\rm B} n_{\rm c}^2 }. 
	\label{strom}
\end{equation} 
When the shadowing effect is remarkable, LyC photons can escape only if a photon never encounter any clumps, and hence the escape fraction is expected to be controlled by the covering factor of clumps.
Even if $R_{\rm c} < l_{\rm s}$ holds and LyC photons completely ionize clumps, the inhomogeneous density field enhances the total recombination rate in a finite volume.\footnote{For instance, it has been well known that the clumpiness of the IGM delays the reionization process \citep{Madau99}. }
In the case that almost entire volume of a galaxy is ionized, the total recombination rate proportional to $C_{\rm f}\langle n_{\rm gas}\rangle^2$ becomes a key quantity for determining the escape fraction, where $C_{\rm f}$ is the clumping factor defined as $\langle n_{\rm gas}^2\rangle / \langle n_{\rm gas}\rangle^2$ and the brackets $\langle \rangle$ denotes the volume weighted average. 
In our calculation, $C_{\rm f}$ is an indirect parameter that increases with $f_{\rm c,m}$.

In contrast to our parameterization, FS11 treated $n_{\rm c}$ as an indirect parameter that is controlled by other direct parameters, i.e., the density contrast between the clumps and inter-clump regions, $C$, and the volume filling factor of clumps, $f_{\rm v}$.
With the parameterization in FS11, $n_{\rm c}$ synchronously decreases as $f_{\rm v}$ increases for a fixed $C$. 
As shown by Equation~(\ref{strom}), this behavior of $n_{\rm c}$ means that shadowing effect of each clump becomes less important with the increasing number of clumps. 
Although the origin of clumps in high-$z$ galaxies is still unknown, there seems to be no affirmative motivation for such a correlation between $n_{\rm c}$ and $f_{\rm v}$.
Furthermore, as discussed later in \S~\ref{sec:clumps}, how the covering fraction by clumps affects the escape fraction is dependent on the number density in a clump. 
Therefore, we treat $n_{\rm c}$ as a direct parameter.
In order to make comparison between our study and FS11 clear, We compare our clump parameters, i.e., $f_{\rm c,m}$ and $n_{\rm c}$, with FS11 clump parameters, i.e., $f_{\rm v}$ and $C$.
In the case with $f_{\rm c,m} = 0.5$, the volume filling factor of clumps, $f_{\rm v}$, is $4.4\times 10^{-3}$. 
Also, the density contrast C increases exponentially with radius since the inter-clump density is given by Equation~(\ref{eq:inter_eq}). 
In the case with $n_{\rm c}=100$ and $Z=0$, the relation is $(r,C) = (r_{\rm h},1.36),\ (5r_{\rm h},110),\ (6r_{\rm h},1790)$.
For the fiducial model, we assume the clumps as massive molecular cloud like structures seen in nearby galaxies, since these properties of clumps in high-$z$ galaxies are still uncertain.

There are some theoretical and observational studies on the properties of clumps in low-z galaxies. 
\citet{2014MNRAS.439..936F} simulated the formation of giant molecular clouds across a galaxy, and found that the typical size and density of the clouds are respectively $\sim 10~{\rm pc}$ and $\sim 1000~{\rm cm}^{-3}$.  
\citet{2017MNRAS.464..491F} observed galaxies at $z=1-3$, and found that the diameters of clumps are ranging from 100 to 800~pc, although small clumps could not be resolved due the limited spatial resolution of 100~pc in their observations.
According to these studies, we employ fiducial parameters of $n_{\rm c}=100~{\rm cm^{-3}}$ and $R_{\rm c}=10~{\rm pc}$ and vary these.
Following the uniform probability, the clumps are randomly distributed across a galaxy. 

For the numerical computation, we use three dimensional cartesian grids whose spatial size is 2~pc to sufficiently resolve the clumpy structures. 
As shown by Figure~\ref{fig:schematic}, we only consider a semicircle of the upper half of a galaxy in order to save the computational cost. 
For a halo with $M_{\rm halo}=10^9~\rm M_{\odot}$, we use $1652\times826$ grids for the directions parallel to the disc plane (i.e., $x$-, and $y$-axes in Figure~\ref{fig:schematic}), and $207$ grids for the vertical direction (i.e., $z$-axis). 
For more massive systems, we increase the total number of the cartesian grids so that the spatial grid size is always fixed to 2~pc.

\subsection{Model of ionizing source}
Since we are curious about LyC emitting galaxies during the EoR, we suppose that LyC photons are emitted from metal-poor Population II stars with the metallicity of $0.02Z_{\odot}$. 
We employ a formulation proposed by \cite{2002A&A...382...28S} to determine the number of LyC photons emitted per unit time by a single star, $\bar{Q}$, which is given by 
\begin{align}
&{\rm log}_{10}[\bar{Q}/{\rm s^{-1}}]\nonumber\\
&=
\begin{cases}
27.80+30.68x-14.80x^2+2.50x^3\ \ \ {\rm for}\ \  m \geq 5M_{\odot},\nonumber\\
0\ \ \ \ \ \ \ \ \ \ \ \ \ \ \ \ \ \ \ \ \ \ \ \ \ \ \ \ \ \ \ \ \ \ \ \ \ \ \ \ \ \ \ \ \ {\rm otherwise},\nonumber
\end{cases}
\end{align}
where  $x\equiv {\rm log}_{10}(m/M_{\odot})$ and the photon number is averaged over the lifetime of a star. 
We adopt a Salpeter initial mass function $f(m) \propto m^{-2.35}$ with the lower and upper mass limits respectively being $m_1 = 0.1M_{\odot}$ and $m_2 = 100M_{\odot}$ \citep{1955ApJ...121..161S}. 
The total number of LyC photons generated in a galaxy per second, $Q_{\rm LyC}$, is given by
\begin{equation}
Q_{\rm LyC}=\frac{\int^{m_2}_{m_1} \bar{Q}(m)f(m){\rm d}m}{\int^{m_2}_{m_1} mf(m){\rm d}m}\times M_{\ast}=3.5\times 10^{46}\frac{M_{\ast}}{M_{\odot}}~\rm s^{-1}.
\end{equation}
In this paper, we assume that each ionizing source (Pop. II star cluster) emits the same amount of LyC photons. 
Thus, the production rate of LyC photons per ionizing source is $Q_{\rm star}=Q_{\rm LyC}/N_\ast$, where $N_\ast$ is the number of ionizing sources; 
the LyC emission rate of each ionizing source decreases as the number of sources increases if $M_\ast$ is fixed. 
We place ionizing sources on the galactic mid-plane (i.e., $Z=0$), assuming that the probability distribution of the sources follows the density profile of the galactic disc.

The escape fraction is expected to decrease as stars evolve \citep{2000ApJ...531..846D}. 
However, we neglect the stellar evolution because our interest in this paper is understanding how the internal structures of galaxies, i.e., distributions of stars and clump, affect the escape fraction.

\subsection{Calculating escape fraction} 
We calculate the balance between the rates of incoming LyC photons and recombination along a light ray by performing ray-tracing from all ionizing sources to all directions with the individual small solid angle of $\Delta \Omega_\ast=1.45\times10^{-6}~$ steradian.
Therefore, the number of LyC photons in the small solid angle $\Delta \Omega_{\ast}$ of a single light ray is given by $Q_{\rm star}\Delta \Omega_{\ast}/4\pi$.
Assuming the ionization equilibrium, the number of LyC photons escaping  along a light ray cast from $i$-th ionizing source to the IGM is given by
\begin{align}
	Q_{i,~\rm esc}(\theta_\ast,\phi_\ast)=\left(\frac{Q_{\rm star}}{4\pi}-\alpha_{\rm B}
	\int_0^{r_{\rm bound}} n_{\rm gas}^2(r,\phi,\theta)r_{\ast}^2{\rm d}r_{\ast}
	\right)\Delta \Omega_{\ast}.
	\label{eq:fesc1}
\end{align}
where $n_{\rm gas}$ is the number density of the galactic gas, $\alpha_{\rm B}$ is the case-B recombination coefficient of hydrogen and $r_{\rm bound}$ is the distance between the source and the computational boundary. 
Since the second term in the right-hand side indicates the number of recombination along a given ray, the equation means the net number of LyC photons along the ray with $\Delta \Omega_{\ast}$.
The coordinate ($r, \phi, \theta$) is originated at the galaxy centre, while the variables with the subscript $\ast$ are originated at the position of the ionizing source. 
While $\alpha_{\rm B}$ depends on the temperature of gas, we now assume a uniform gas temperature of $10^4\ {\rm K}$ for which $\alpha_{\rm B} = 2.6\times 10^{-13}\ {\rm cm}^3\ {\rm s}^{-1}$. 
If a value obtained by Eq.~(\ref{eq:fesc1}) results in negative, we set $Q_{i,~\rm esc}(\theta_\ast,\phi_\ast)=0$ which means that no LyC photons can escape in this direction.

Since we employ multiple ionizing sources in contrast to FS11, ionized regions often overlap with each other. 
In this case, we frequently overestimate the total recombination rate if we use Eq.~(\ref{eq:fesc1}) for all ionizing sources without any recipes. 
To avoid the unphysical overestimate, we do NOT count the recombination events at the grid points ionized by another ionizing source already. 
In addition, we never solve Eq.~(\ref{eq:fesc1}) up to $r_\ast=r_{\rm bound}$ at once for each ionizing source, but solve it little by little while changing ionizing source in operation so that the resultant escape fraction hardly depends on the order of calculation (see Appendix B). 
The number of escaping LyC photons for each ionizing source can be evaluated by integrating Eq.~(\ref{eq:fesc1}) over the entire solid angle. 
Then the summation for all ionizing sources gives the total number of LyC photons escaping from a galaxy. 
Thus, the total escape fraction, $f_{\rm esc}$, is found by
 \begin{equation}
	f_{\rm esc}=\frac{1}{Q_{\rm LyC}}\sum_{\rm star}\sum_{\rm angle}Q_{i,\rm esc}(\theta_\ast,\phi_\ast)
\label{eq:fesc}
 \end{equation}

When we calculate the escape fraction, we simultaneously derive the covering factor of clumps. 
For this purpose, we calculate the fraction of sky covered by clumps as seen from each ionizing source, and take the average of the fractions over all ionizing sources. 
With this definition, the covering factor $f_{\rm cover}$ ranges from 0 to 1. 

As described above, we only consider a representative one-fourth region of the whole structure to save the numerical cost. 
Even with such an incomplete computational domain, we can take into account the comings and goings of LyC photons between the computational domain and other virtual domains by letting the rays reflect at the boundaries, assuming symmetric distributions of ionizing sources and clomps. 
The detailed way of the reflection and some tests for verifying our method are shown in Appendixes A and C.

It should be noted that we neglect dust and helium in our calculations. 
Neglecting helium atoms would lead to decrease in the hydrogen LyC escape fraction, because the number of hydrogen atoms increases for a fixed total gas mass. 
It has also been known that photo-heating of helium increases the temperature of ionized gas and slightly affect the recombination rate, though its impact is remarkable only if radiation spectrum is hard \citep{Kitayama01}. 
The roles of dust concerning the escape fraction though the absorption and scattering processes are too complicated to be considered in our simple modeling. 
Although the quantitative impacts of dust are still under debate \cite[e.g.][]{2008ApJ...672..765G,2010ApJ...710.1239R,2011MNRAS.412..411Y}, dust has been expected to work to reduce the escape fraction. 
Despite the fact that such simplification actually prevents us from predicting quantitative values of the escape fraction, our simple modeling can help us to understand the behavior of the escape fraction. 
The aim in this paper is to clarify the relation between the internal structures of galaxies and the resultant escape fraction.

%
%

\section{RESULTS \& DISCUSSIONS}\label{sec:RESULTS}
In this section, we present our numerical results regarding the dependence of the escape fraction on various quantities of galaxies. 
Although we are especially curious about influence of clumps, we first show the results for the smooth gas disc model with no clumps, because this relatively simple model helps our understanding.  
 
\subsection{Behavior of the escape fraction in smooth gas disc model}\label{sec:smooth}
We first present the dependence of the escape fraction on the number of ionizing sources $N_{\ast}$, the LyC emission rate of each source, and halo mass that have not been explored by FS11. 
Throughout this subsection, we employ the smooth gas disc model with $f_\ast=0.1$. 

\subsubsection{Effects of multiple ionizing sources}\label{pos}
\begin{figure}
\includegraphics[width=\columnwidth]{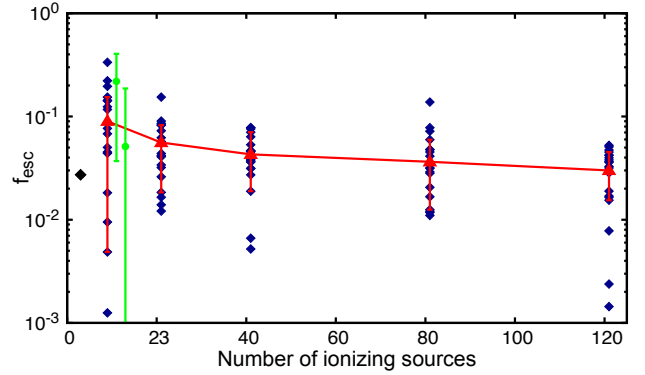}
	\caption{Escape fraction as a function of the number of ionizing sources for the 
	smooth disc model with $M_{\rm halo}=10^9~\rm M_{\odot}$, 
	$\langle v^2\rangle=GM_{\rm halo}/r_{\rm vir}$ (i.e., type-A galaxy) and $f_*=0.1$. 
	We perform 20 runs for each $N_\ast$ with different disposition of ionizing 
	sources, 
	and denote the resultant escape fraction for each run by a blue diamond. 
	Error bars represent the 70 percentiles of 20 resultants.
	The arithmetic mean value of the escape fractions (red circles)
	declines with increasing the number of ionizing sources. 
	The filled circles are two examples with $N_{\ast}=9$ and the associated error bars correspond to the variance originated in viewing angle (see Section \ref{sightline})}
	The filled diamond mark represents escape fraction for a single ionizing source at the galactic center.
	\label{fig:number_star}
\end{figure}
Figure~\ref{fig:number_star} shows the escape fractions in the case  of $M_{\rm halo}=10^9~\rm M_{\odot}$ for various number of ionizing sources. 
For the calculation, we assume $\langle v^2\rangle=GM_{\rm halo}/r_{\rm vir}$ corresponding to the disc scale hight $z_0=(GM_{\rm halo}/2\pi G \rho_0 r_{\rm vir})^{0.5}$. 
Since we here adopt $f_*=0.1$ and $m_{\rm d}=0.155$, the mass and LyC emission rate of each ionizing source respectively correspond to $M_*/N_* = 1.55\times10^7/N_*~\rm M_{\odot}$ and $5.4\times 10^{53}/N_*~{\rm s^{-1}}$. 
For each $N_\ast$, we perform 20 runs with different disposition of ionizing sources to evaluate the scatter and average values of the escape fraction. 
From Figure~\ref{fig:number_star}, we can see the following two remarkable trends:  
(i) the scatter becomes smaller as $N_*$ increases, and (ii) the average value decreases slightly with $N_*$. 
The first trend is straightforwardly expected from the statistical point of view. 
When $N_*\sim10$, the escape fraction varies $\sim 3$ orders of magnitude depending on the disposition of ionizing sources, while the dispersion in the case with $N_*\sim120$ is almost equal to an order of magnitude. 
The second trend may be statistically unreliable but can be understood as follows; the number of LyC photons per an ionizing source increases by decreasing the number of ionizing sources. 
Hence, with less number of ionizing sources, each ionizing source can create a large \ion{H}{ii} region which makes it easy for LyC photons to escape.   
In the opposite case, the typical bubble size around an ionizing source is relatively small, and LyC photons can escape only when \ion{H}{ii} regions overlap each other to create a larger \ion{H}{ii} region. 

We confirm that this trend is maintained as far as we consider generally expected range of $f_\ast=0.01-0.3$, though the scatter becomes smaller for larger $f_{\ast}$. 
It simply comes from the fact that LyC photons emitted from sources can escape regardless of their positions if $f_\ast$ is high. 
We also confirm that this argument is valid even for halos with $M_{\rm halo}>10^9\rm M_\odot$.

Our findings give us an important caveat in numerical studies of the escape fraction; the mean escape fractions derived from numerical simulations may depend on the mass resolution. 
In fact, it has been known that  the escape fraction seems to increase on average as the mass resolution becomes worse \cite[e.g.][]{Wise14,Paardekooper15,Kimm17}. 
\citet{Wise14} speculated that the mean escape fraction in low resolution simulations tends to be high because low mass resolution generally results in instantaneous birth of numerous stars and hence SFR becomes very high at this moment. 
Our findings indicate that the increase in the escape fraction is also caused by the individual size of \ion{H}{ii} regions even if SFR is fixed.

\subsubsection{Dependence on halo mass for smooth disc model}\label{mdepend_s}
\begin{figure}
	\includegraphics[width=\columnwidth]{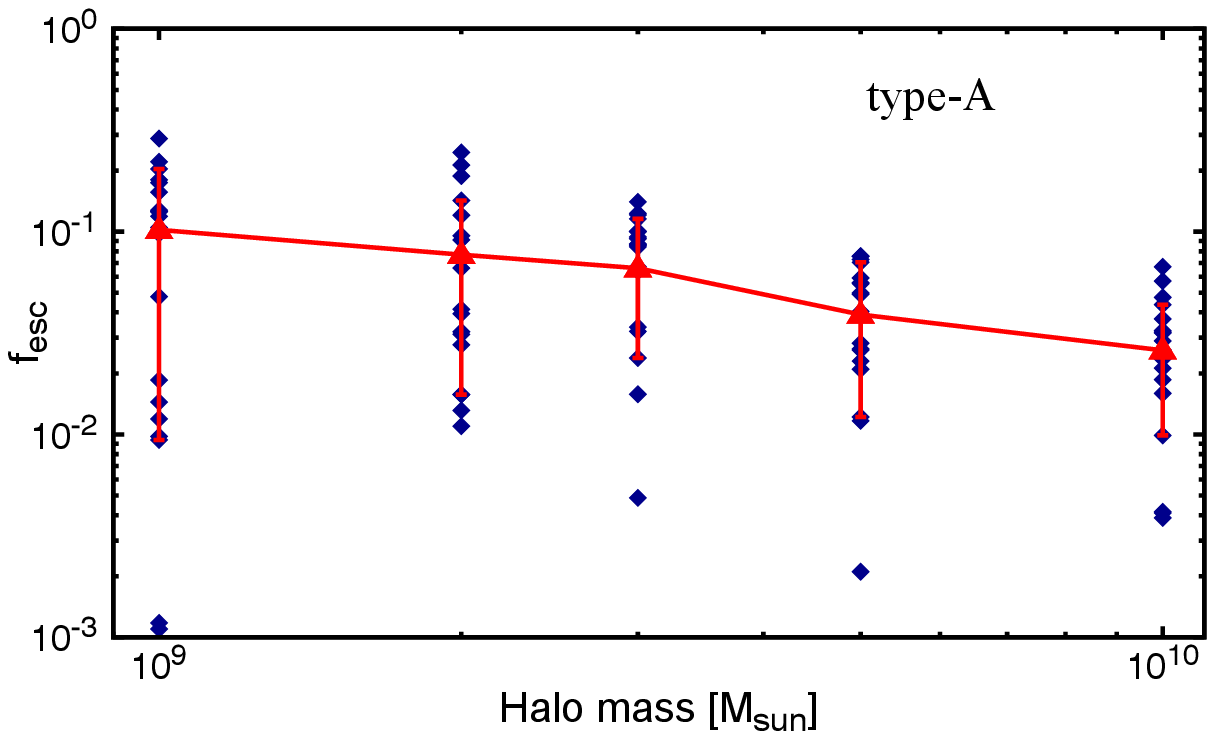}
	\includegraphics[width=\columnwidth]{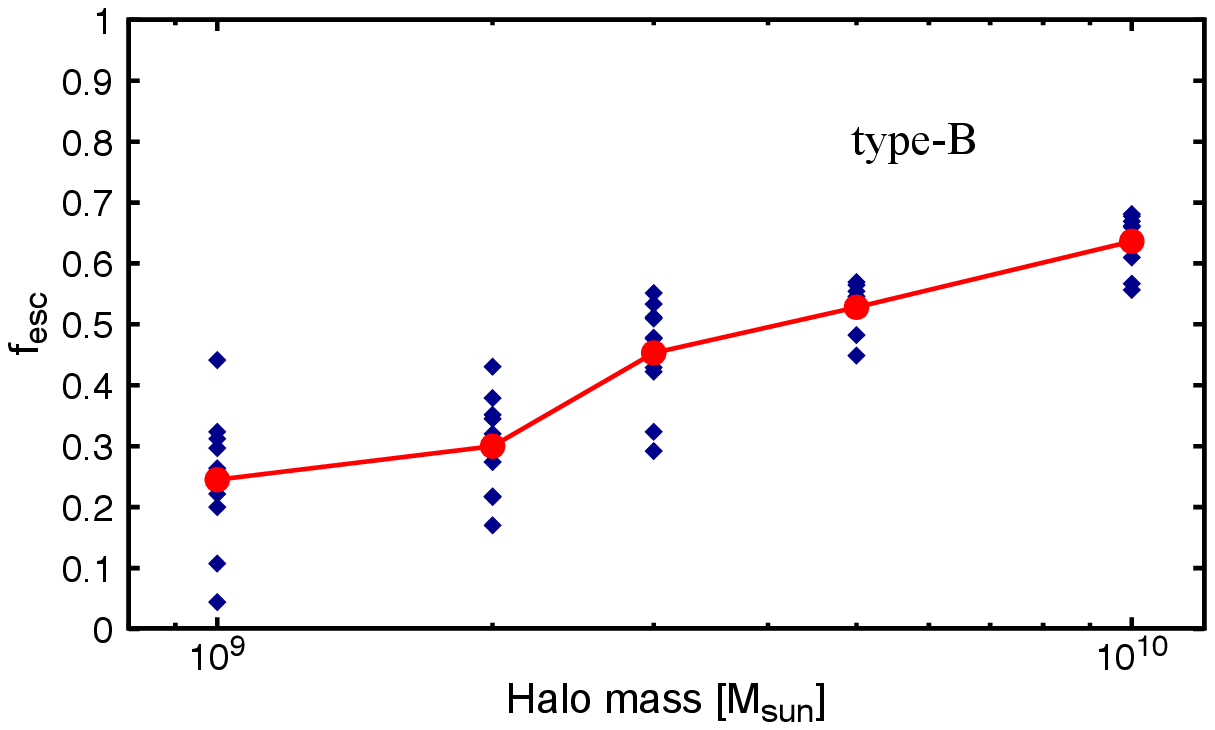}
	\caption{Escape fraction as a function of halo mass for the smooth disc model 
	with $N_{*}=9\times(M_{\rm halo}/{\rm 10^9~M_{\odot}})$, and $f_*=0.1$. 
	The upper and bottom panels show the results for type-A galaxies 
	($\langle v^2\rangle=GM_{\rm halo}/r_{\rm vir}$) and type-B galaxies ($\langle v^2\rangle = 10~{\rm km~s^{-1}}$), respectively. 
	We perform 20 runs for each $N_\ast$ with different disposition of ionizing 
	sources, and denote the resultant escape fraction for each run by a blue diamond.  
	In the upper panel, the arithmetic mean value of the escape fractions (red circles)
	declines with increasing halo mass, while in the lower panel, the average escape fractions positively 
	correlate with halo mass. }
	\label{fig:halo_star}
\end{figure}
We elucidate how the escape fraction depends on halo mass by employing type-A and type-B galaxies. 
For both cases, we assume the stellar mass fraction of $f_\ast=0.1$, and set the number of ionizing sources to be $N_* = 9\times(M_{\rm halo}/{\rm 10^9~M_{\odot}})$ so that the mass of an ionizing source is identical regardless of halo mass. 
Here, one of the nine ionizing sources is always placed at the galactic centre, and the others are placed in accordance with the gas density profile of the smooth disc. 

The upper panel of Figure~\ref{fig:halo_star} shows the escape fraction as a function of halo mass for type-A galaxies. 
We obtain the mass dependence of the escape fraction similar to the majority of previous studies \cite[e.g.][]{2010ApJ...710.1239R,2011MNRAS.412..411Y,Wise14,Paardekooper15,Kimm17}; 
the mean escape fraction decreases with increasing halo mass. 
Complicated properties of simulated galaxies often make it difficult to understand this trend, while our simple model brings a brief understanding of the trend. 
Since we assume a discy structure, LyC photons escape more easily in the direction perpendicular to the galactic plane, rather than the parallel directions. 
For a type-A galaxy, both the scale radius and the scale height are proportional to $M_{\rm halo}^{1/3}$ while the central hydrogen number density $n_0$ is independent of halo mass (red lines in Figure~\ref{fig:galala}). 
Therefore, the recombination events per unit time along the perpendicular direction can be roughly estimated as $\sim \alpha_{\rm B}n_{0}^2z_{0}^3\propto M_{\rm halo}$. 
Since LyC emission rate per an ionizing source is independent of halo mass, this estimate indicates that LyC photons tend to be unlikely to escape from a galaxy as halo mass increases, as shown by Eq.~(\ref{eq:fesc1}). 
\footnote{The escape fraction would be independent of halo mass if we consider only one ionizing source whose emission rate is proportional to halo mass. }
It is worth mentioning that similar dependence on halo mass can be obtained even for round-shape  galaxies by replacing $z_0$ with the radius of a round-shape galaxy. 
The upper panel of Figure~\ref{fig:halo_star} also shows that the scatters in $f_{\rm esc}$ are as large as $\sim1$-$3$ orders of magnitude. 
Here we emphasize that these large scatters are produced solely by the disposition of ionizing sources. 
Although the scatters shown here seem to be a bit smaller than those shown by previous state-of-the-art simulations utilizing a large sample of simulated galaxies \cite[e.g][]{2011MNRAS.412..411Y, Kimm17}, the simulated galaxies with given mass have a variety of SFRs which should cause an additional scatter in $f_{\rm esc}$. 

The bottom panel of Figure~\ref{fig:halo_star} is similar to the upper one but for type-B galaxies. 
We find that the escape fraction increases on average with halo mass. 
This tendency is similar to that in \cite{2008ApJ...672..765G}. 
Let us make a rough estimate again for this case. 
In contrast to the aforementioned case, the central hydrogen number density is not constant but proportional to $M_{\rm halo}^{2/3}$, and thus $n_0^2z_0^3 \propto M_{\rm halo}^{1/3}$ (blue lines in Figure~\ref{fig:galala}). 
Since the total number of LyC photons is proportional to halo mass, the escape of LyC photons becomes easier with increasing halo mass. 
We note that the absolute value of the escape fraction is expected to be sensitive to the scale height, but the dependence on halo mass shown here hardly alters as far as the thermal velocity dispersion $\langle v^2\rangle$ is independent of halo mass. 

From Figure~\ref{fig:halo_star},  we expect that the galaxy shape affects the halo mass dependence of the escape fraction. 
In fact, galaxies in previous simulations showing an anticorrelation between $f_{\rm esc}$ and $M_{\rm halo}$ seem to be irregular or round-shaped, while galaxies in a simulation showing the opposite trend have a prominent gaseous disc \cite[see Figure~8 in][]{2008ApJ...672..765G} although the dependence of the scale height on halo mass is not clear for us. 
Thus, we argue that the galaxy shape is a cause of the different dependence of the escape fraction on halo mass seen in previous simulations \cite[e.g.][]{2008ApJ...672..765G,2010ApJ...710.1239R,2011MNRAS.412..411Y,Wise14,Paardekooper15,Kimm17}. 
As expected from this argument, accurately simulating morphology of galaxies is likely crucial to understand the halo mass dependence of the escape fraction, although some inconsistency still remains even among state-of-the-art simulations of an isolated galaxy \citep{AGORA}. 

\subsubsection{Sightline dependence}\label{sightline}
\begin{figure}
	\includegraphics[width=\columnwidth]{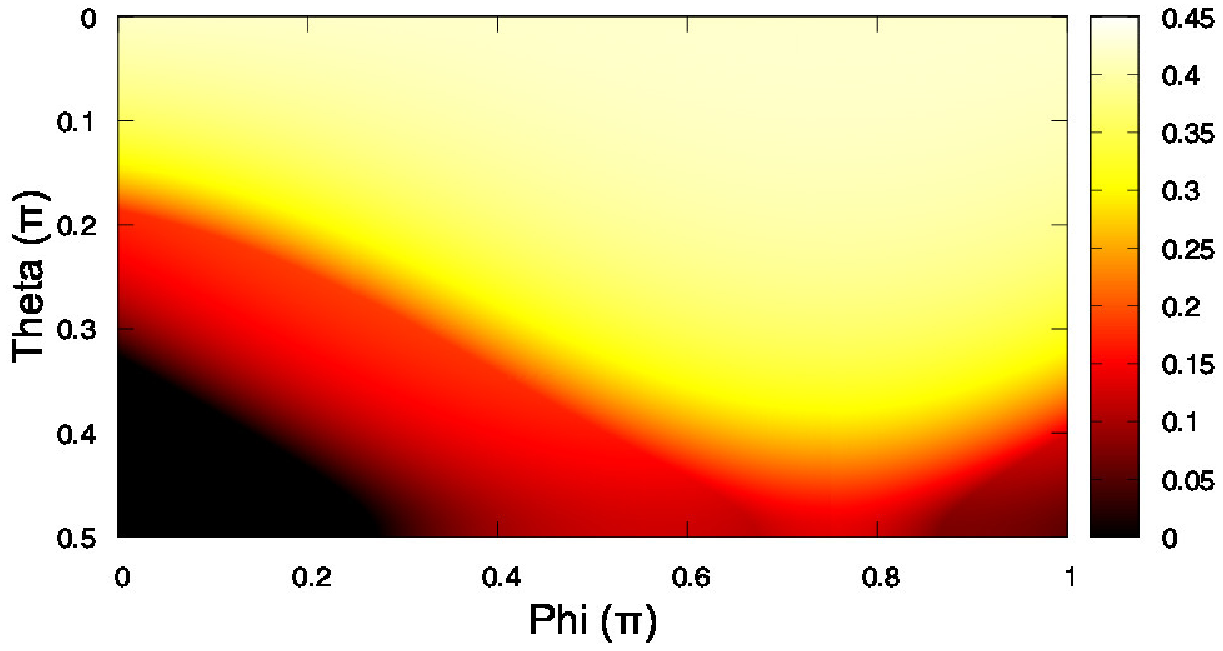}
	\includegraphics[width=\columnwidth]{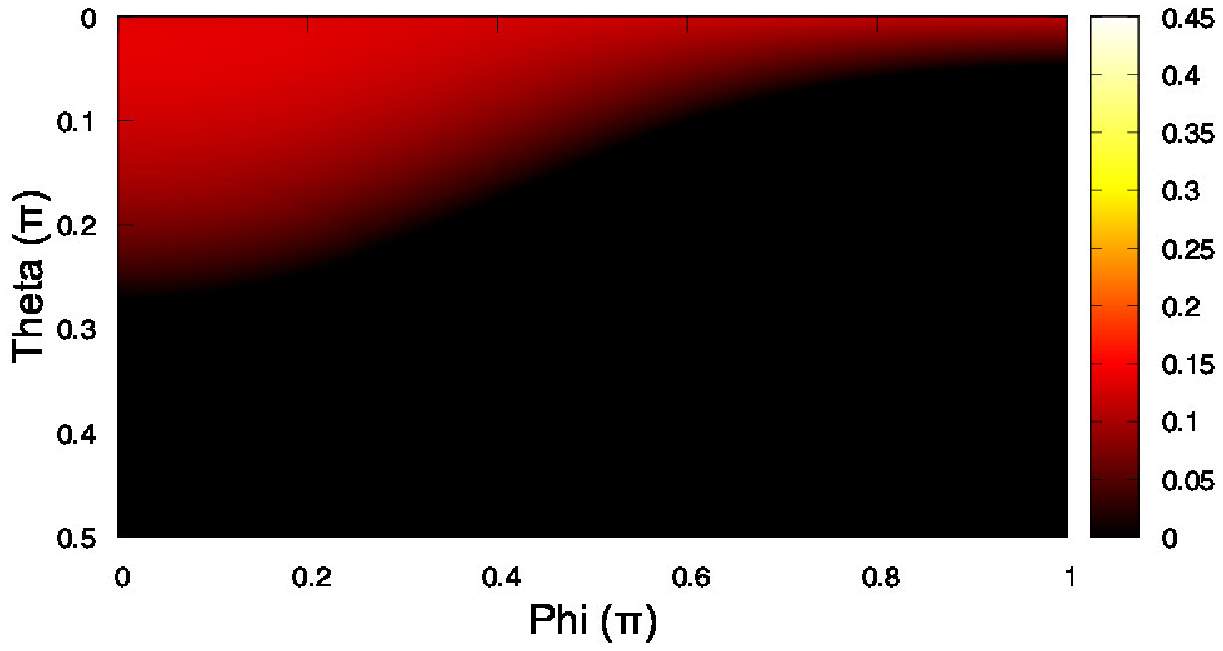}
	\caption{Dependence of the escape fraction on viewing angle. 
	We employ type-A galaxy model with $M_{\rm halo}=10^9~\rm M_{\odot}$, $f_*=0.1$, 
	and $N_\ast=9$. 
	The upper and lower panels respectively shows the results with different disposition of ionizing 
	sources.   
	$M_{\rm halo}=10^9~\rm M_{\odot}$ and $f_*=0.1$ but different mean escape fractions. 
	The color represents the value of the escape fraction. 
	The upper panel shows the angular distribution of the escape fraction for a galaxy with
	$f_{\rm esc} = 0.24$.
        The lower panel shows that for a galaxy with 
        $f_{\rm esc} = 0.02$.	
	}
	\label{fig:angle}
\end{figure}

The total escape fraction is important to evaluate the contribution to reionization from galaxies. 
However, observers measure the escape fraction in a specific line-of-sight and it also causes the diversity in the observed escape fraction.  
Therefore, we show the sightline-dependence of the escape fraction in this section.
We consider the smooth disc type-A model with $M_{\rm halo}=10^9~\rm M_{\odot}$, $f_*=0.1$, and $N_\ast=9$ to evaluate the impact of the disposition of ionizing sources on the anisotropy of the escape fraction. 

Figure~\ref{fig:angle} shows that the escape fraction as a two dimensional function of azimuthal direction, $\phi$, and the angle between the $z$-axis and a sightline, $\theta$ (see Figure~\ref{fig:schematic}). 
Since the disposition of ionizing sources causes large diversity in the mean escape fraction, we show the sightline-dependent escape fractions for two galaxies with high mean escape fraction of 0.21 (the upper panel of Figure~\ref{fig:angle}) and low  escape fraction of 0.05 (the lower panel of Figure~\ref{fig:angle}).  
We should firstly note that the escape fraction significantly changes with viewing angle for both galaxies. 
As shown by Figure~\ref{fig:schematic}, the scatter caused by viewing angle is sometimes larger than that caused by the placements of ionizing sources.
We confirm that this trend holds even in cases with $N_{\ast}=81$ because the scatter caused by viewing-angle weakly depend on N*. 
Though it may be unexpected, the escape fraction slightly depends on $\phi$ direction because we employ only nine ionizing sources. 
More notable feature is the dependence on $\theta$. 
As mentioned in \S~\ref{mdepend_s}, LyC photons escape more easily in the polar direction ($\theta\sim0$). 
In the both galaxies, the escape fractions in $\theta<0.2\pi$-$0.3\pi$ directions are relatively high. 
There seems to be a critical angle above which the escape fraction sharply drops, though the value of the critical angle seems to depend on the disposition of sources, disc structure, and so on.
This indicates that we sometimes miss LyC photons from galaxies. 

%
%
\subsection{Effects of clumpy gas distribution on escape fraction}\label{sec:clumps}
\subsubsection{Diversity of escape fractions caused by disposition of ionizing sources in the clumpy gas disc.}\label{sec:scatter_c}
\begin{figure}
	\includegraphics[width=\columnwidth]{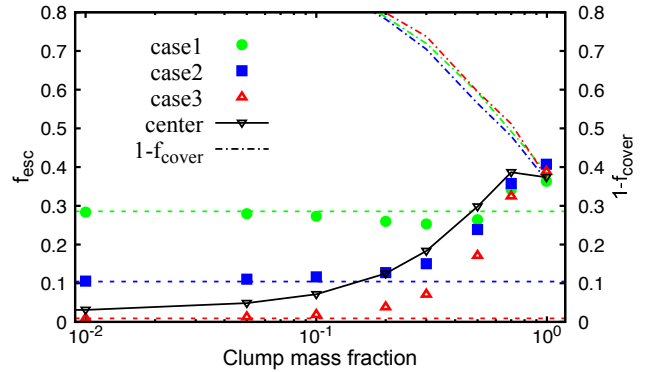}
	\caption{Escape fraction as a function of the total clump mass fraction $f_{\rm c,m}$ for type-A 
	clumpy disc galaxies with $M_{\rm halo}=10^9~\rm M_{\odot}$, 
	$n_{\rm c}=100~{\rm cm^{-3}}$, $R_{\rm c}=10~{\rm pc}$ and $f_\ast=0.1$. 
	We perform 10 runs for each $f_{\rm c,m}$ with different placementes of clumps. 
	Red, blue and green denote the resultant escape fraction for each ionizing source distribution. 
	The dashed and dot-dashed lines show the escape fractions in the smooth disc model and $1-f_{\rm cover}$, respectively.
	The scatter caused by the disposition of ionizing sources becomes smaller  
	with increasing the clump mass fraction.}
	\label{fig:scatter_clump}
\end{figure}

\begin{table*}
\begin{center}
\begin{tabular}{ccccccc}
\hline
Case &  disposition of ionizing source ($r_{\ast}$(pc),$\phi_{\ast}({\rm rad})$) & $f_{\rm esc,\ast}$ (no clump) & $f_{\rm esc,\ast}$ in $f_{\rm c,m} = 0.1$ &  $f_{\rm esc,\ast}$ in $f_{\rm c,m} \simeq 1.0$ & $f_{\rm cover,\ast}$ in $f_{\rm c,m} \simeq 1.0$\\
\hline
1      & (0,0)   & 0 & 0 & $4.26 \times 10^{-2}$  & $7.87 \times 10^{-1}$ \\
        & (188,2.14)   &$ 1.46 \times 10^{-3}$ & $3.62 \times 10^{-3}$ & $4.00 \times 10^{-2}$ &$5.84 \times 10^{-1}$ \\
        & (97,0.565)   & 0 & 0 & $4.15 \times 10^{-2}$ & $6.00 \times 10^{-1}$\\
        & (490,2.96)   & $8.11 \times 10^{-2}$ & $7.44 \times 10^{-2}$ & $3.70 \times 10^{-2}$& $6.26 \times 10^{-1}$\\
        & (423,1.19)  & $6.03 \times 10^{-2}$ & $5.85 \times 10^{-2}$ & $4.19 \times 10^{-2}$& $5.50 \times 10^{-1}$\\
\hline
2    & (0,0)  & 0  & 0 & $4.37 \times 10^{-2}$ & $7.87 \times 10^{-1}$ \cr
      & (90,0.613)  & 0  & 0 & $4.41 \times 10^{-2}$ & $6.06 \times 10^{-1}$ \cr
      & (74,3.14)  & $9.15 \times 10^{-6}$   & $6.93 \times 10^{-5}$ & $4.52 \times 10^{-2}$ & $6.12\times 10^{-1}$ \cr
      & (256,0.478)  & 0 & $2.49 \times 10^{-2}$ & $4.54 \times 10^{-2}$  & $6.18 \times 10^{-1}$\cr
      & (224,0)  &  $4.58 \times 10^{-3}$ & $ 3.31 \times 10^{-2}$ & $4.71 \times 10^{-2}$ &$6.20 \times 10^{-1}$ \cr
\hline
3    & (0,0)  & 0  & 0 & $4.41 \times 10^{-2}$ & $7.87 \times 10^{-1}$ \cr
      & (160,1.50)  & 0  & $4.36 \times 10^{-4}$ & $4.24 \times 10^{-2}$ & $5.27 \times 10^{-1}$ \cr
      & (145,1.21)  & $3.95 \times 10^{-6}$   & $8.07 \times 10^{-4}$ & $4.29 \times 10^{-2}$ & $6.84\times 10^{-1}$ \cr
      & (56,0.510)  & $2.10 \times 10^{-2}$ & 0 & $4.48 \times 10^{-2}$  & $6.20 \times 10^{-1}$\cr
      & (207,2.78)  &  $3.10 \times 10^{-2}$ & $ 8.91 \times 10^{-3}$ & $4.27 \times 10^{-2}$ &$5.93 \times 10^{-1}$ \cr
\hline
\end{tabular}
\caption{ The disposition of ionizing sources, $f_{{\rm esc},\ast}$ of case 1, case 2 and case 3 for Figure~\ref{fig:scatter_clump}.
$f_{{\rm esc},\ast}$ means the escape rate of escaping LyC photon emitted by each ionizing source to emitted LyC photons by all ionizing sources.
$f_{{\rm cover},\ast}$ also means the covering factor of a ionizing source.
In this case, scale radius becomes $r_{\rm h}= 165\ ({\rm pc})$.
Note: the four other ionizing sources are placed symmetrically in the "virtual" domain.
}
\label{tab:case}
\end{center}
\end{table*}


We first show the dependence of the escape fraction on the clump mass fraction, $f_{\rm c,m}$, and then present the scatter caused by the dispositions of ionizing sources and clumps after that. 
To estimate the diversity caused by the disposition of ionizing sources, we employ three cases of the dispositions, hereafter Case1, Case 2, and Case 3. 
The positions of the sources in these cases are summarized in Table~\ref{tab:case}. 
The parameters are set to be $N_* = 9$, $n_{\rm c}=100~{\rm cm^{-3}}$, $R_{\rm c}=10~{\rm pc}$ and $f_\ast=0.1$. 

Figure~\ref{fig:scatter_clump} shows the escape fractions for these three cases as a function of the clump mass fraction, $f_{\rm c,m}$. 
We find that a critical clump mass fraction of $\sim 10^{-1}$ below which the escape fractions are almost identical to those for the smooth disc model (dashed lines in Figure~\ref{fig:scatter_clump}). 
Therefore, as similar to the smooth disc model, the escape fraction is controlled by the positions of ionizing sources in the low $f_{\rm c,m}$ regime. 
In contrast, in the high $f_{\rm c,m}$ regime, the escape fraction increases with $f_{\rm c,m}$. 
To understand this behavior, we also plot $1-f_{\rm cover}$, where $f_{\rm cover}$ is the covering fraction by clump, which is defined as the average of those looked at from each ionizing source.
As shown in Figure~\ref{fig:scatter_clump}, $f_{\rm cover}$ is too small to prevent LyC photons from escaping even if $0.1<f_{\rm c,m}<0.5$. 
Therefore, the decrease in the density of inter-clump region essentially determines the dependence of $f_{\rm esc}$ on $f_{\rm c,m}$ at $0.1<f_{\rm c,m}<0.5$. 
Interestingly, in the high clump mass fraction regime ($f_{\rm clump}\sim 1$), the escape fraction accords with $1 - f_{\rm cover}$.

We further discuss the scatter caused by the dispositions of ionizing sources and clumps.
In Case 1, ionizing sources are distributed in outer regions, where gas density is small. 
As a result, Case 1 shows higher escape fraction than those in other two cases when $f_{\rm c,m}$ is small. 
It is worth mentioning that there is almost no difference in the escape fraction of three cases in the high clump mass fraction regime ($f_{\rm c,m}> 0.7$) because  these cases show the same value of covering factor.
In this regime, the inter-clump gas density is enough low to hardly contributes the LyC absorption, and the view from an ionizing source is almost independent of its position since the clumps are uniformly distributed across the galaxy.

In order to investigate the scatter caused by the placements of clumps, we performed 10 runs by fixing the parameter set but varying the placements of clumps. 
As a result, we find that the scatter caused by the placements of clumps is significantly small. 
In the low clump mass fraction regime ($f_{\rm c,m}<0.1$), clumps hardly affect the escape fraction as shown by Figure~\ref{fig:scatter_clump}. 
The scatter in the high clump mass fraction regime ($f_{\rm c,m}\sim1$) is also small because of abundant clumps. 
In the intermediate clump mass fraction regimes ($f_{\rm c,m}\sim0.5$), the scatter is slightly larger than the above extreme cases, but is still smaller than the size of symbols in Figure~\ref{fig:scatter_clump}. 
The scatter shown in FS11 is relatively larger than ours. 
The difference is likely originated in the different parameterization of clumps as mentioned in \S~\ref{sec:Galactic Geometry}.

Lastly, we compare our results for multiple ionizing sources to that for a single big ionizing source at the galactic center, which is shown by inverse triangles in Figure~\ref{fig:scatter_clump}. 
In the low $f_{\rm c,m}$ regime, we find that the escape fraction for the central single source case is smaller than that of Case1 and 2 because of high density around the galactic center. 
On the other hand, the escape fraction of the central source case is larger than that of Case 3 which multiple sources reside around the galactic center, because the single luminous ionizing sources can create a large ${\rm H}_{\rm II}$ region.
Interestingly, in the high clump mass fraction regime ($f_{\rm c,m}\sim1$), the escape fraction of center case is the same value as that of Case 1, 2 and 3.

\begin{figure}
	\includegraphics[width=\columnwidth]{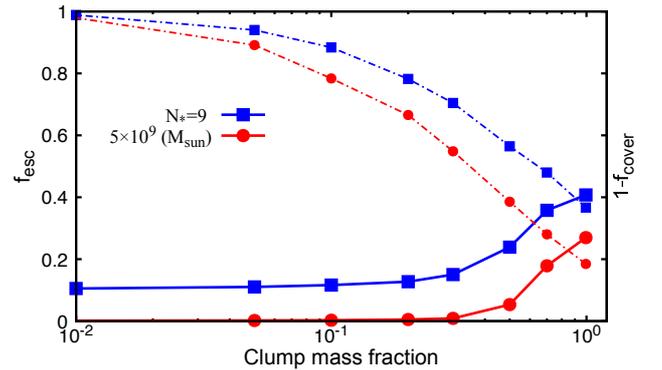}
	\caption{
	Escape fraction as a function of the total clump mass fraction $f_{\rm c,m}$ for the 
	clumpy disc model with $N_* = 9\times(M_{\rm halo}/{\rm 10^9~M_{\odot}})$, 
	$n_{\rm c}=100~{\rm cm^{-3}}$, $R_{\rm c}=10~{\rm pc}$ and $f_\ast=0.1$. 
	The blue and red solid lines denote the mean resultant escape fractions 
	$M_{\rm halo}=10^9~\rm M_{\odot}$ and $M_{\rm halo}=5\times10^9~\rm M_{\odot}$, 
	respectively. 	
	The dot-dashed lines show $1-f_{\rm cover}$ for each case. 
	}
	\label{fig:nstarmass_clump}
\end{figure}

\subsubsection{Dependence on halo mass in clumpy disc model}\label{}
In Figure~\ref{fig:nstarmass_clump}, we show the dependence of the escape fraction on the clump mass fraction for galaxies with two different masses of  $M_{\rm halo}=10^9~\rm M_{\odot}$ and $M_{\rm halo}=5\times10^9~\rm M_{\odot}$. 
Relevant parameters are set to be $N_* = 9\times(M_{\rm halo}/{\rm 10^9~M_{\odot}})$, $n_{\rm c}=100~\rm cm^{-3}$, $R_{\rm c}=10~\rm pc$, and $f_\ast=0.1$.
As similar to the result for the smooth disc model (\S~\ref{mdepend_s}), the escape fraction decreases with increasing halo mass as far as the properties of clumps are independent of halo mass.
In the case of low $f_{\rm c,m}$, the reason for this mass dependence is the same as the case of the smooth disc model, because the clumps do not work in this regime. 
We also find that the dependence on halo mass does not alter even if $f_{\rm c,m}$ is very high, because the covering factor of clumps increases with halo mass as far as the properties of clumps are fixed. 

The derived halo mass dependence of the escape fraction is similar to those presented by high resolution simulations \cite[e.g.,][]{Wise14,Paardekooper15, Kimm17}. 
The anticorrelation between $f_{\rm esc}$ and $M_{\rm halo}$ is one of the reasons why less massive galaxies are expected to be responsible for cosmic reionization.
However, the escape fractions derived from simulations with sufficient spatial resolution are hitherto limited to halos with $M_{\rm halo}\lesssim10^9~\rm M_\odot$, and the escape fractions for massive galaxies still remain unclear. 
As shown in \S~\ref{mdepend_s}, the halo mass dependence is likely sensitive to the galaxy shapes. 
Therefore, understanding the relation between the galactic mass and morphology would be important not only for quantifying the contribution from galaxies to cosmic reionization but also for revealing the galaxy evolution.

\subsubsection{Dependence on size and gas density of clumps}\label{sec:prop_clump}
\begin{figure}
	\includegraphics[width=\columnwidth]{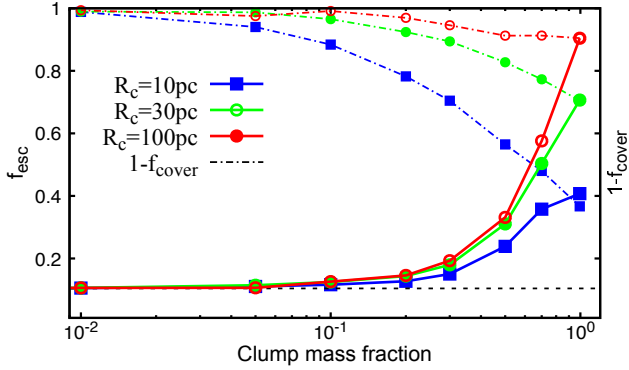}
	\caption{Escape fraction as a 
	function of $f_{\rm c,m}$ for $R_{\rm c}=10~\rm pc$ (filled squares), 
	$R_{\rm c}=30~\rm pc$ (filled circles) 
	and $R_{\rm c}=100~\rm pc$ (open circles). 
	The other parameters are set to be $M_{\rm halo}=10^9~\rm M_{\odot}$, 
	$N_\ast=9$, $n_{\rm c}=100~{\rm cm^{-3}}$, and $f_\ast=0.1$. 
	The dashed and dot-dashed lines indicate the arithmetic mean escape fraction 
	for the smooth disc model and $1-f_{\rm cover}$, respectively. 
	The escape fraction becomes higher as $R_{\rm c}$ becomes larger 
	because of lower covering factors for larger $R_{\rm c}$. 
	When the total clump mass dominates the disc mass (i.e. $f_{\rm c,m}\gtrsim0.5$), 
	the escape fractions are almost equal to $1-f_{\rm cover}$ if the size of clumps is large.}
	\label{fig:m9_cr}
\end{figure}
In this subsection, we further examine the dependence of the escape fraction on the size of clumps.
Assuming an ionizing source at the centre of a galaxy and equating $R_{\rm c}$ to  $l_{\rm s}$ derived from Eq.~(\ref{strom}), we obtain the critical distance as 
\begin{equation}
	r_{\rm cr} \sim 1.5r_{\rm h}\times\left(\frac{f_\ast}{0.1}\right)^{\frac{1}{2}}
	\left(\frac{10~{\rm pc}}{R_{\rm c}}\right)^{\frac{1}{2}}
	\left(\frac{100~{\rm cm^{-3}}}{n_{\rm c}}\right)
	\left(\frac{M_{\rm halo}}{10^9~{\rm M_\odot}}\right)^{-\frac{1}{3}}. 
	\label{eq:rcr}
\end{equation}
Roughly speaking, clumps reside inside $1.5r_{\rm cr}$ will be fully ionized, thus can not block LyC photons with the parameters assumed in this section. 
When the value of the critical distance is large, the shadowing effect does not work anymore.
Thus, it is important to examine the dependence of the escape fraction on the size of clumps. 
Here we assume $M_{\rm halo}=10^9~\rm M_\odot$, $N_\ast=9$, $n_{\rm c}=100~\rm cm^{-3}$, and $f_\ast=0.1$. 
In Figure~\ref{fig:m9_cr}, we compare the escape fractions for three clump diameters of 10 pc, 30 pc and 100 pc. 
The escape fraction increases with increasing $R_{\rm c}$ regardless of the clump mass fraction. 
Equation~(\ref{eq:rcr}) tells us that the importance of the shadowing effect on the escape fraction becomes more crucial with increasing $R_{\rm c}$.
The number of clumps $N_{\rm c}$ and a covered area per one clump $A_{\rm c}$ respectively follow $N_{\rm c} \propto R_{\rm c}^{-3}$ and $A_{\rm c}\propto{R_{\rm c}^{2}}$ for a fixed clump mass fraction. 
Hence the covering factor is approximately proportional to $N_{\rm c} \times A_{\rm c} \propto R_{\rm c}^{-1}$, neglecting overlap along the line of sight. 
In fact, Figure~\ref{fig:m9_cr} shows that the covering factor becomes higher with decreasing $R_{\rm c}$. 
It should be noted that the escape fraction almost obeys $1-f_{\rm cover}$ when the total clump mass dominates in the disc and the size of clumps is large.

\begin{figure}
	\includegraphics[width=\columnwidth]{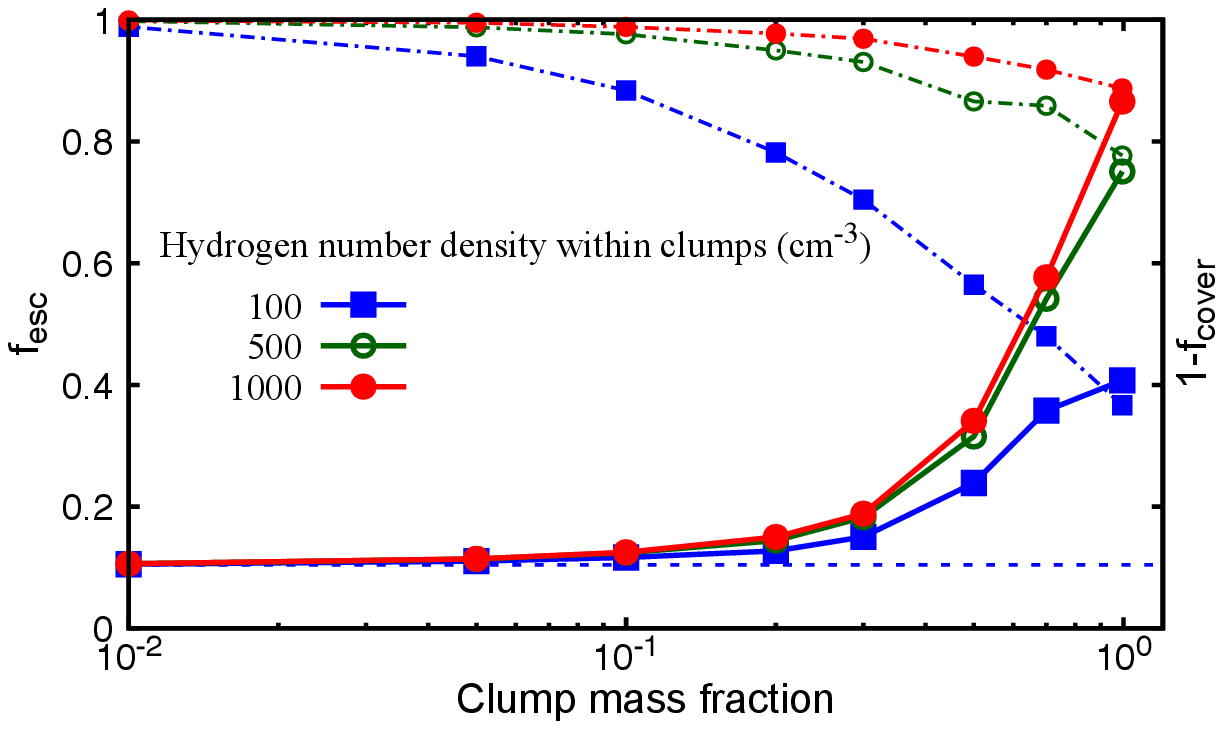}
	\includegraphics[width=\columnwidth]{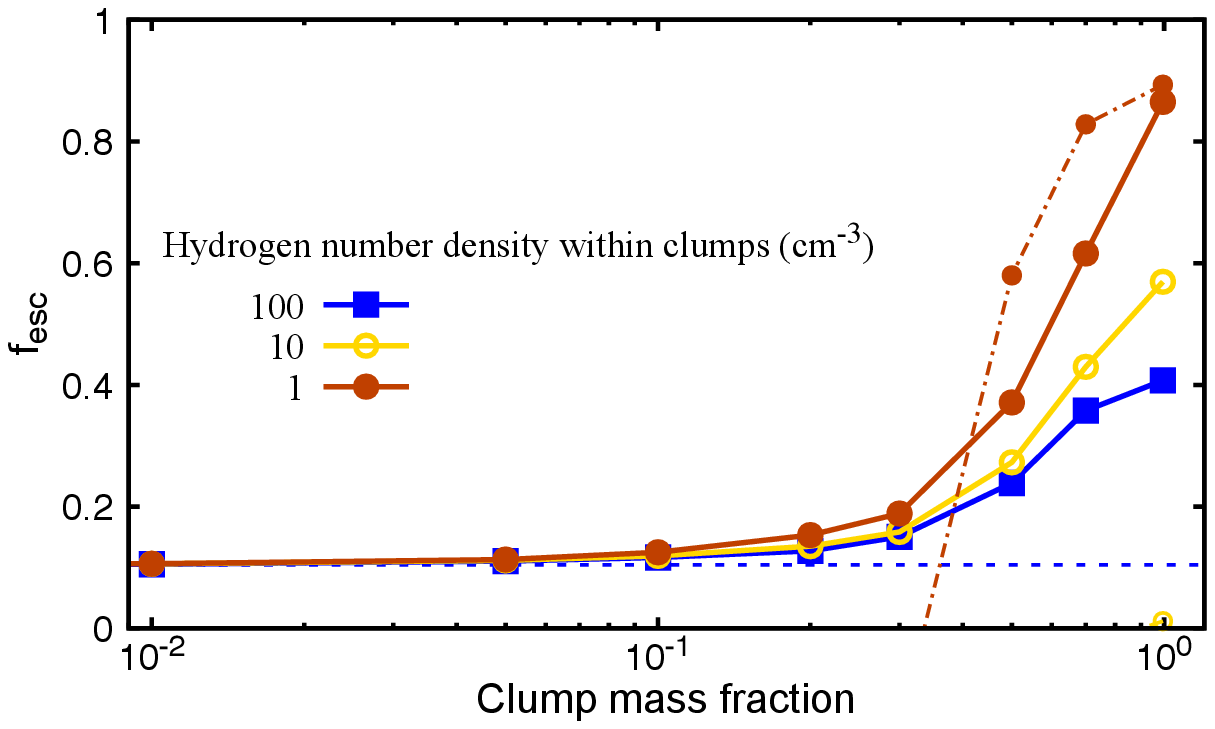}
	\caption{Escape fraction as a function of $f_{\rm c,m}$ for various values of 
	the internal number density of clumps $n_{\rm c}$. 
	The other parameters are set to be $M_{\rm halo}=10^9~\rm M_{\odot}$, 
	$N_\ast=9$, $R_{\rm c}=10~{\rm pc}$, and $f_\ast=0.1$. 
	{\it Upper panel}: the escape fractions for $n_{\rm c}=100~\rm cm^{-3}$, 
	$500~\rm cm^{-3}$, and $1000~\rm cm^{-3}$ are indicated by 
	the filled square, open square, and filled circle symbols, respectively. 
	The dot-dashed lines correspond to $1-f_{\rm cover}$ that are well consistent with 
	the computed escape fraction at $f_{\rm c,m}\gtrsim0.5$. 
	{\it Lower panel}: the escape fractions for $n_{\rm c}=1~\rm cm^{-3}$, 
	$10~\rm cm^{-3}$, and $100~\rm cm^{-3}$ are indicated by 
	the cross, open triangle, and filled square symbols, respectively. 
	The dot-dashed lines correspond to Eq.~(\ref{eq:fesc2}) (see text.). 
	In the case of extremely small clump density ($n_{\rm c}=1~\rm cm^{-3}$), 
	the escape fractions at $f_{\rm c,m}\gtrsim0.5$ are almost equal to Eq.~(\ref{eq:fesc2}).}
\label{fig:m9_n}
\end{figure}
Next, we show in Figure~\ref{fig:m9_n} the effects of the internal gas density of clumps on $f_{\rm esc}$ for the fixed clump size of $R_{\rm c}=10~\rm pc$. 
The upper and lower panels in Figure~\ref{fig:m9_n} show the results for high density clumps ($n_{\rm c}\geq100~\rm cm^{-3}$) and for low density clumps ($n_{\rm c}\leq100~\rm cm^{-3}$), respectively. 
The dependence of the escape fraction on $n_{\rm c}$ in the two cases are quite different; 
For high density clumps, the escape fraction increases with increasing $n_{\rm c}$, while for low density clumps, the lowest $n_{\rm c}$ model shows the highest escape fraction. 
As shown in Eq.~(\ref{eq:rcr}), the escape fraction is controlled by the shadowing effect if $n_{\rm c}$ is higher than $100~\rm cm^{-3}$. 
Since the increase in $n_{\rm c}$ leads to the decrease in the covering factor as $f_{\rm cover}\propto n_{\rm c}^{-1}$,  the escape fraction increases as $n_{\rm c}$ increases. 
The escape fractions for high $n_{\rm c}$ cases can be approximated as $f_{\rm esc}\approx1-f_{\rm cover}$ at $f_{\rm c,m}\approx1$ as already shown.  

On the other hand, Eq.~(\ref{eq:rcr}) indicates that the shadowing effect is unlikely to work anymore if $n_{\rm c}\leq 10~\rm cm^{-3}$. 
Therefore, the escape fraction does not decrease despite the increase of the covering factor. 
When almost entire volume of the disc is ionized, we expect that the escape fraction is roughly expressed as 
\begin{equation}
	f_{\rm esc} \approx \frac{Q_{\rm LyC}-C_{\rm f}\alpha_{\rm B}\langle n_{\rm gas}\rangle^2 V_{\rm gal}}{Q_{\rm LyC}},
	\label{eq:fesc2}
\end{equation}
where $V_{\rm gal}$ is the total volume of the galaxy. 
We show in the lower panel of Figure~\ref{fig:m9_n} that the escape fractions for $n_{\rm c}=1~\rm cm^{-3}$ are roughly in line with Eq.~(\ref{eq:fesc2}) in the high $f_{\rm c,m}$ regime. 
To conclude, the escape fractions for low density clumps are controlled by the clumping factor $C_{\rm f}$.  

Our results suggest that knowledge of the properties of clumps is important to understand the escaping process of LyC photons. 
Although a correlation between star formation activity and the escape fraction is often discussed \cite[e.g.][]{Wise14,Kimm17}, it is also important to focus on a relationship between the inhomogeneity caused by such star formation activity and the escape fraction \citep{Umemura12}. 
Our results also imply that the spatial resolution of numerical simulations likely affects the resultant escape fraction. 
It is generally true that simulations with low spatial resolution fail to resolve small scale structures sufficiently. 
The lack of resolution results in the appearance of artificially diffuse structures and thus in the overestimate of $f_{\rm esc}$ as implied from our results. 
Therefore high spatial resolution is leastwise necessary for predicting reliable values of the escape fraction. 

%
%

\section{SUMMARY}\label{sec:SUMMARY}
In this paper, we have investigated how the properties of galaxies affect the escape fraction. 
For the purpose, we have employed two simple disc galaxy models, i.e, the smooth and clumpy disc models, and performed ray-tracing calculation through the galaxies to derive the escape fractions. 

First we have explored the influence of multiple ionizing sources on the scatter and average values of LyC escape fractions with a smooth disc model. 
As a result, we have found that less number of ionizing sources leads to a larger scatter up to $\sim 3$ orders of magnitude and to a higher escape fraction. 
The former trend is due to the sensitive dependence of the escape fraction on the disposition of ionizing sources, while the latter can be explained by the increase in the production rate of LyC photons per ionizing source. 
We have also explored the dependence of the escape fraction on halo mass for the smooth disc model, and found that the halo mass dependence is determined by the behavior of the scale height of the gas disc. 
The escape fraction decreases as halo mass increases if the scale height is determined by the virial temperature and thus depending on halo mass as $z_0\propto M_{\rm halo}^{1/3}$. 
Therefore, the recombination rate integrated along the direction perpendicular to the disc plane proportionally increases with halo mass. 
In contrast, the escape fraction positively correlates with halo mass if the scale height is independent of halo mass due to relatively weak dependence of the integrated recombination rate on halo mass. 
We have also investigated the sightline dependence, which is expected when we observe leaking LyC photons. 
We have shown that the diversity caused by the sightline is sometimes larger than that caused by the disposition of ionizing sources. 
The scatter is likely large in the case that the total escape fraction is small, because LyC photons can escape along a limited angle.

Next we have elucidated the impacts of clumps in the gas disc on the escape fraction. 
As a result, we have found that the clumpy structures hardly affect the escape fraction as far as the total clump mass fraction is smaller than $\sim10^{-1}$. 
At the regime of $10^{-1} \lesssim f_{\rm c,m}$, the escape fraction basically increases with the clump mass fraction, because LyC photons can escape through gaps among the clumps. 
The resultant escape fraction reaches up to $\sim 5$ times higher than that for the smooth disc case, depending on the properties of clumps. 
If clumps are dense and/or large enough to obstruct LyC photons, the escape fraction is determined by the covering factor of the clumps $f_{\rm cover}$ and obeys $f_{\rm esc}\approx1-f_{\rm cover}$. 
In the case with low density clumps, the clumps are almost ionized and the escape fraction is controlled by the clumping factor as shown by Eq.~(\ref{eq:fesc2}). 
We have also investigated the halo mass dependence of the escape fraction for the clumpy disc model. 
As a result, we have found that the halo mass dependence is qualitatively the same as that for the smooth disc model as far as the properties of clumps are independent of halo mass. 
To conclude, the properties of clumps and the galaxy shape play crucial roles in determining the escape fraction and its halo mass dependence. 
Our model used for evaluating the escape fraction may be too simple to predict reliable values of the escape fraction.
However our results certainly tell us important quantities for determining the escape fraction as well as how they regulate the escape fraction. 
Our results manifest that quantifying the properties of galaxies (i.e., the typical density, size, and amount of clumps as well as entire morphology of galaxies) and clarifying their halo mass dependence are essential for understanding the escaping process of LyC photons. 

\section*{Acknowledgements}
We are grateful to Kiyotomo Ichiki, Akio K. Inoue, and Masayuki Umemura for comments on our results that improved our study. 
This work was supported in part by the Grant-in-Aid for Scientic Research on Innovative 15H05890 (T. S.), the Grant-in-Aid for JSPS Research Fellows No. 26-3216
(D. K.) and a grant from NAOJ (K. H.). 
We also acknowledge the Kobayashi-Maskawa Institute for the Origin of Particles and the Universe, Nagoya University, for providing useful computing resources for conducting this research.




\bibliographystyle{mnras}
\bibliography{refs} 



\appendix

\section{Ray Reflection Technique}\label{sec:RAY_RE}
\begin{figure}
	\includegraphics[width=\columnwidth]{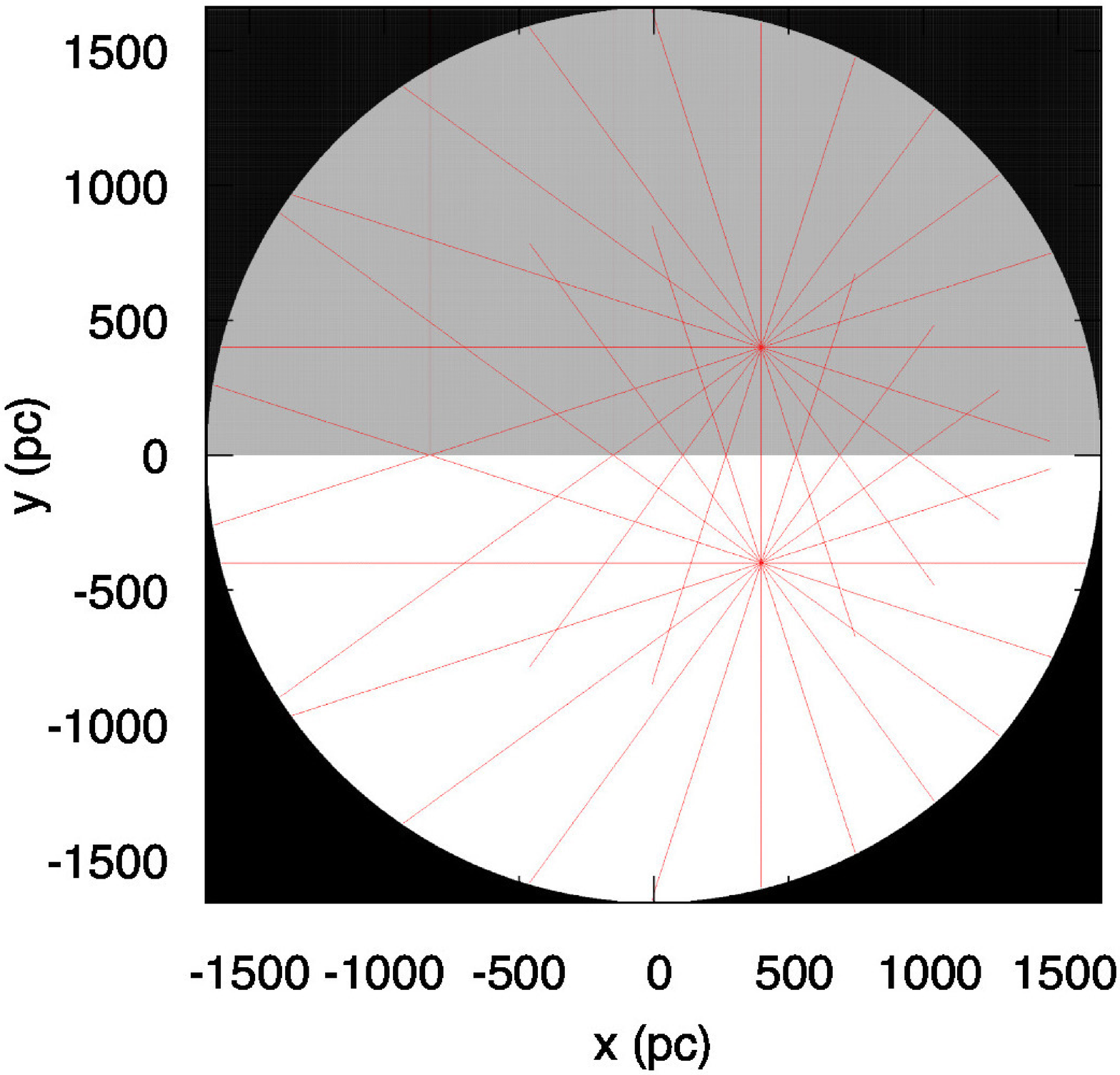}
	\includegraphics[width=\columnwidth]{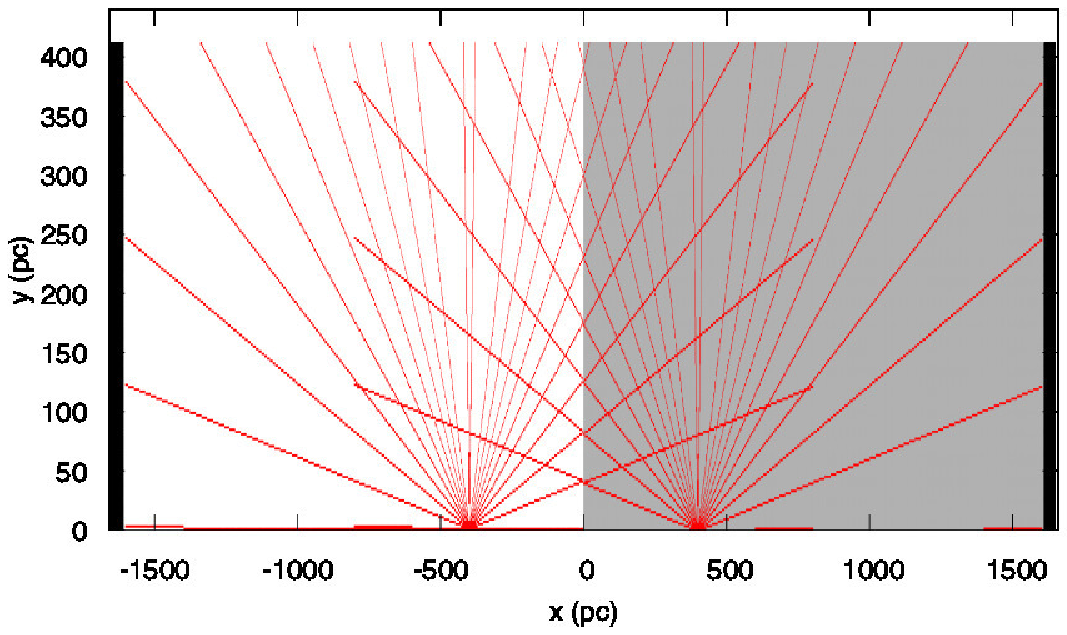}
	\caption{Rays from ionizing sources with the ray reflection technique. 
	The upper and lower panels show the rays on the $x-y$ plane and the $x-z$ plane, respectively.  
	The gray and blank regions respectively corresponds to the computational domain 
	and the virtual domain. 
	The panels nicely demonstrate that we appropriately consider rays from a source 
	in the virtual domain by using the ray reflection technique. 
	}
	\label{fig:xy_test_ray}
\end{figure}

As mentioned in Section~\ref{sec:Galactic Geometry}, we only consider a semicircle of the upper half of a galaxy to save the calculation cost.
We then assume symmetric structures for both clumps and ionizing sources.  
To perform ray tracing in such a partial representative area and derive the {\it total} escape fraction, we reflect rays at the boundaries of the computational domain. 
The upper panel of Figure~\ref{fig:xy_test_ray} demonstrates the ray reflection technique on the $x-y$ plane with 20 rays and the lower panel shows that on the $x-z$ plane with 10 rays. 
We find that rays from an ionizing source in the virtual domain can be appropriately considered with this method. 

\section{Convergence of the computed escape fraction}\label{sec:test}
\begin{figure}
	\includegraphics[width=\columnwidth]{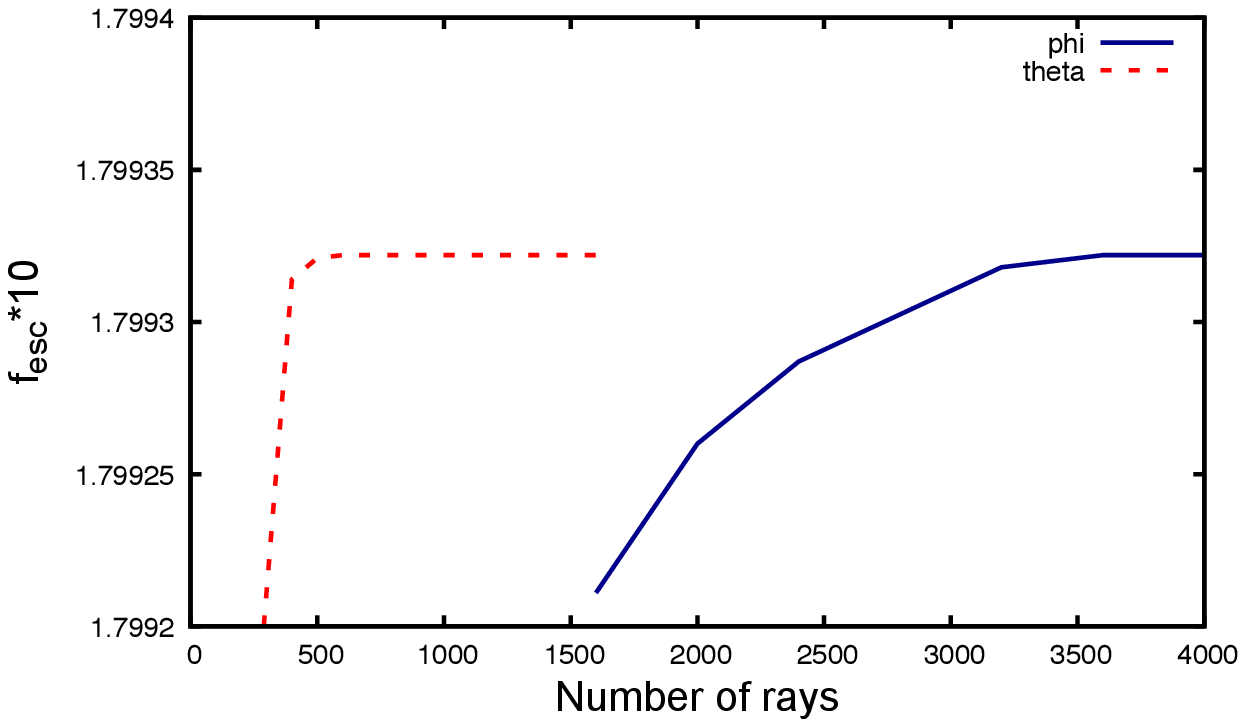}
	\caption{Computed escape fraction as a function of the number of rays casted for the ray-tracing.
	The red dashed curve indicates the number of rays for $\theta$ (angle between $z$-axis and a ray), 
	while the blue solid  curve indicates that for $\phi$ (angle between $x$-axis and a ray). 
	The computed escape fraction converges to a certain value 
	if we use 500 rays across $0\le \cos\theta\le 1$ and 3,600 rays across 
	$0\le \phi < 2\pi$. }
	\label{fig:ap_one}
\end{figure}
\begin{figure}
	\includegraphics[width=\columnwidth]{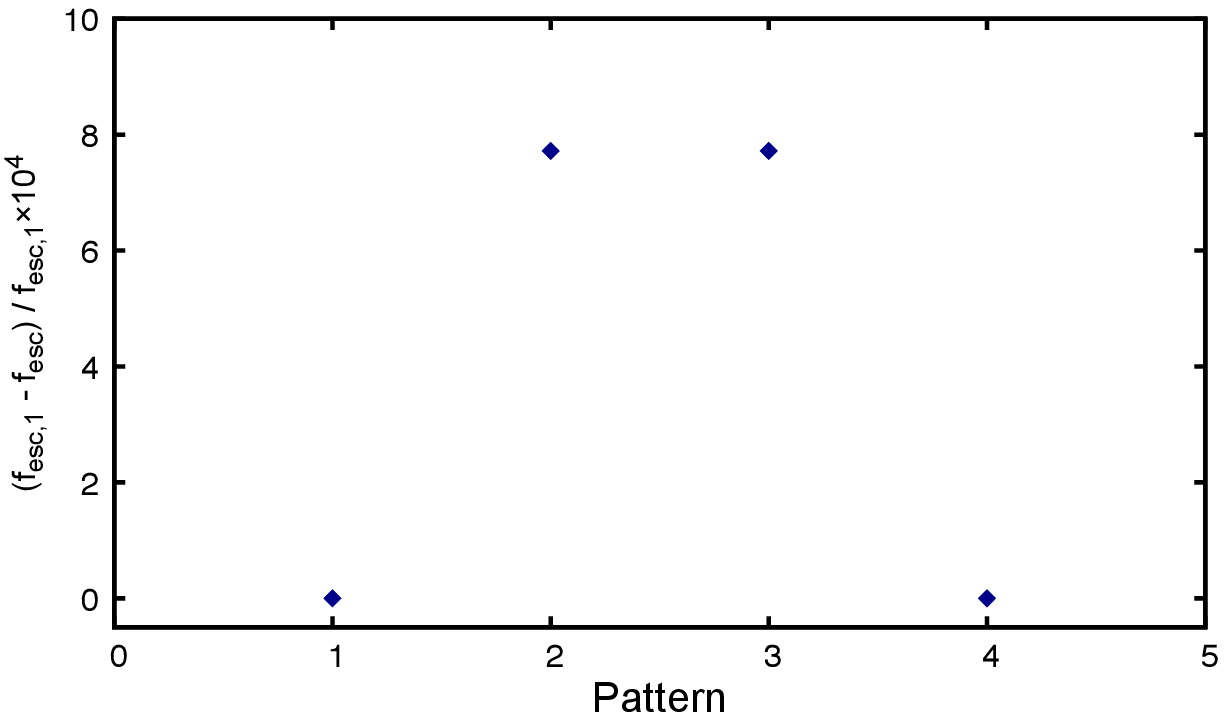}
	\caption{Relative errors in the escape fraction for the various orders of calculation of 
	ionizing sources. We use 9 sources for this test and plot the computed escape
	fractions for some cases in which the orders of calculation of ionizing sources are different 
	from each other. 
	It is found that the relative errors in the resultant escape fraction are less than $10^{-3}$, so that
	the resultant escape fraction hardly depends on the order of calculation. }
	\label{fig:integ}
\end{figure}

Here, we show the validness of our method used in this paper. 
Firstly, Figure~\ref{fig:ap_one} shows the resultant escape fraction as a function of the number of rays used in our calculation. 
We find that 500 rays across $0\le \cos\theta \le 1$ and 3,600 rays across $0\le \phi < 2\pi$ are at least required for the convergence. 
In conclusion, the number of rays employed in our calculations is sufficient to evaluate the escape fraction. 

We also demonstrate in Figure~\ref{fig:integ} that the spatial integration along a ray in our calculation is valid for evaluating the escape fraction. 
In the case of multiple ionizing sources, the evaluated escape fraction depends on the order of calculation of ionizing sources if our method is invalid. 
Figure~\ref{fig:integ} shows that the relative errors in the computed escape fraction are less than $10^{-3}$. 

\section{Computing ionized bubbles}\label{sec:bubble}
\begin{figure}
	\includegraphics[width=\columnwidth]{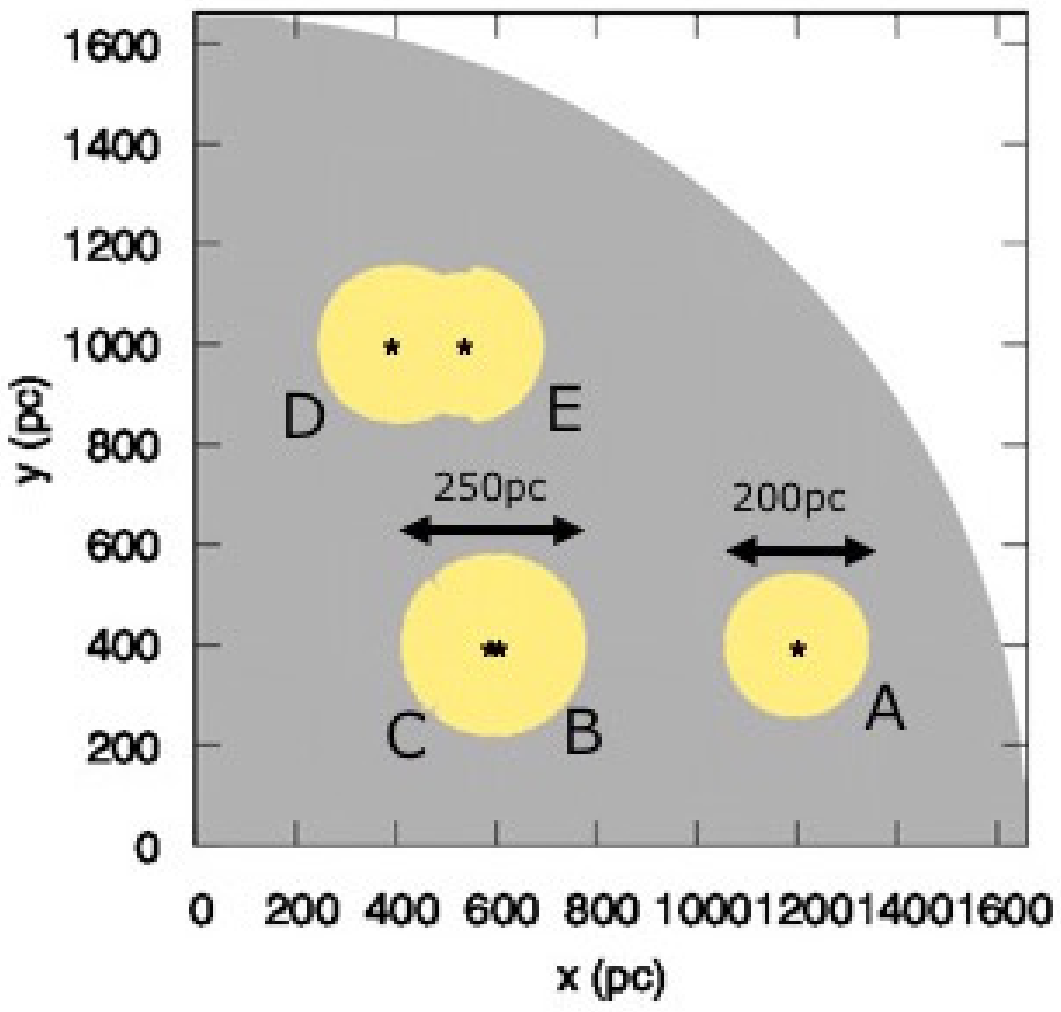}
	\caption{Test for reproducing ionized bubbles. The yellow regions represent ionization regions.}
	\label{fig:over}
\end{figure}

We show whether the sizes of isolated bubbles and overlapped bubbles are calculated correctly or not in Figure~\ref{fig:over}. 
For the test, we assume a uniform disc with $n_{\rm H}=50~\rm cm^{-3}$, and place five ionizing sources with 
$N_{\rm ion}=7.8\times10^{52} {\rm s}^{-1}$ at $(x,y) = (1200 {\rm pc},400 {\rm pc})$ (Source A), $(600 {\rm pc},400 {\rm pc})$ (source B), $(590 {\rm pc},400 {\rm pc})$ (Source C), $(400 {\rm pc},1000 {\rm pc})$ (Source D), and $(540 {\rm pc},1000 {\rm pc})$ (Source E).
With this setup, $\rm Str\ddot{o}mgren$ radius for a single source theoretically corresponds to $r_{\rm s} \approx 100\ {\rm pc}$. 
Figure~\ref{fig:over} shows that the size of ionized region around Source A, an isolated source, well corresponds to the analytical solution. 
Since Sources B and C are closely located, the resultant ionized region roughly corresponds to $\sqrt{2}r_{\rm s}$ as shown by Figure~\ref{fig:over}. 
Finally, the separation between Sources D and E is less than $r_{\rm s}$. Therefore, each ionized region is partially overlapped. 
Although the shape of the overlapped bubble is slightly awkward due to the finite spatial resolution, the resultant volume of the ionized bubble is well consistent with the analytical prediction. 
In summary, our method appropriately reproduce the ionized bubble sizes expected from the analytic solution.

\section{Dependence on stellar mass fraction and clump mass fraction}\label{sec:fstar}
\begin{figure}
	\includegraphics[width=\columnwidth]{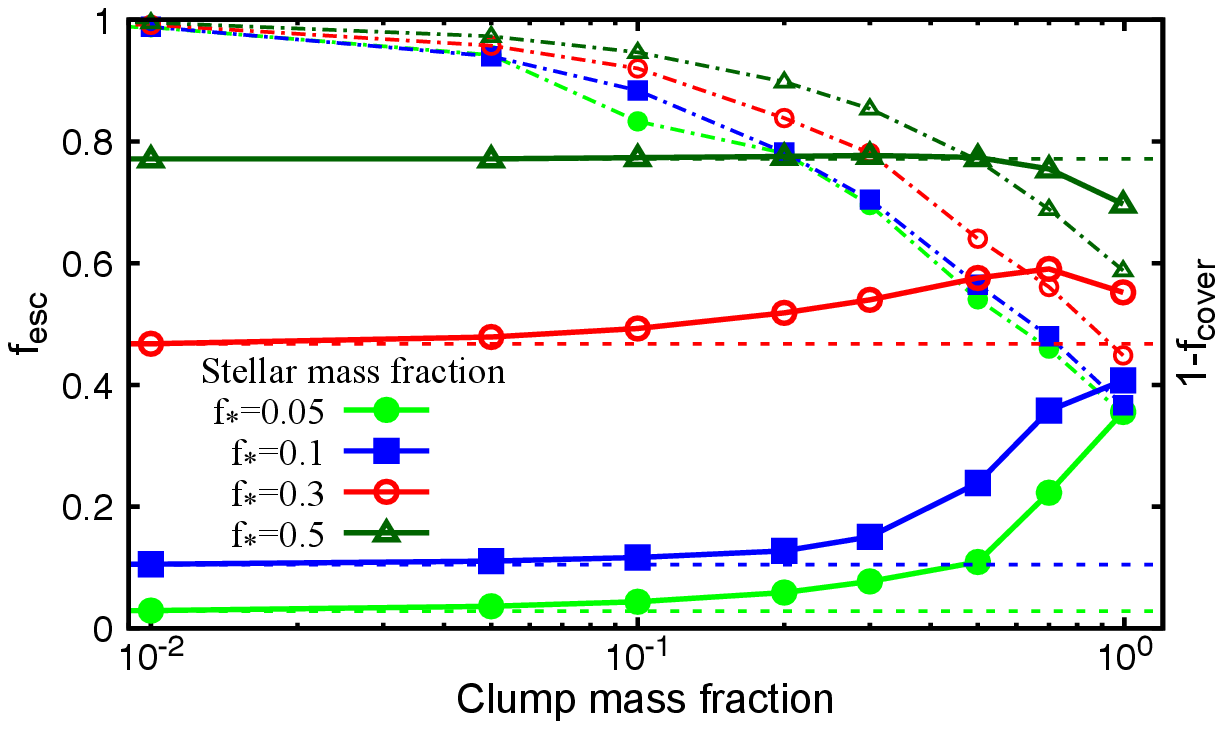}
	\caption{Escape fraction as a function of the clump mass fraction $f_{\rm c,m}$ for various 
	stellar mass fractions $f_{\ast}$. 
	We employ the clumpy disc model with $M_{\rm halo}=10^9~\rm M_{\odot}$, $N_\ast=9$,
	$n_{\rm c}=100~{\rm cm^{-3}}$, and $R_{\rm c}=10~{\rm pc}$. 
	The filled circle, filled square, open circle, and open triangle symbols 
	respectively 
	indicate the results for $f_{\ast}=0.05$, 0.1, 0.3, and 0.5. 
	The dashed and dot-dashed lines show the mean escape fractions in the smooth 
	disc model and $1-f_{\rm cover}$ for each $f_{\ast}$, respectively. 
	The escape fractions begin to divert from those for the smooth 
	disc model at $f_{\rm c,m}\gtrsim 10^{-2}$.
	Except for the high stellar mass fraction model ($f_{\ast}=0.5$), clumpy structure leads to 
	the increase of the escape fraction. 
	In the model with $f_{\ast}=0.5$, the enhancement of clumping factor is remarkable and 
	the escape fraction turns out to be decreased.}
	\label{fig:fstar}
\end{figure}
We show in Figure~\ref{fig:fstar} the escape fraction as a function of the clump mass fraction $f_{\rm c,m}$ for various values of the stellar mass fraction $f_{\ast}$.  
The density and size of the clumps are set to be $n_{\rm c}=100~{\rm cm^{-3}}$ and $R_{\rm c}=10~{\rm pc}$, respectively. 
As shown in the upper panel of Figure~\ref{fig:fstar}, the escape fraction increases with increasing stellar mass fraction for a given $f_{\rm c,m}$ because the number of LyC photons proportionally increases with $f_{\ast}$ while the gas mass in the disc decreases as $M_{\rm gas}=(1-f_{\ast})M_{\rm disc}$. 
Except for the case with $f_{\ast}=0.5$, the escape fractions are higher than those for the the smooth disc model at $f_{\rm c,m}>10^{-2}$, because the hydrogen number density in the inter-clump region begins to decrease. 
As discussed in \S~\ref{sec:prop_clump}, the existence of clumps enhances the recombination rate and decreases the escape fraction when the large fraction of the disc is ionized. 
We can see this effect in the case of high stellar mass fraction with $f_{\ast}=0.5$. 


\bsp	
\label{lastpage}
\end{document}